\documentclass[journal]{IEEEtran}
\usepackage{amsmath,amsfonts,amssymb}
\usepackage{xspace}
\usepackage{algpseudocode}
\usepackage{algorithm}
\usepackage{array}
\usepackage[caption=false,font=normalsize,labelfont=sf,textfont=sf]{subfig}
\usepackage{textcomp}
\usepackage{stfloats}
\usepackage{url}
\usepackage{verbatim}
\usepackage{graphicx}
\usepackage{cite}
\usepackage{color}
\usepackage{hyperref}
\usepackage{booktabs} 
\usepackage{tabularray}
\usepackage{multirow}
\usepackage[normalem]{ulem}
\usepackage{xcolor}
\usepackage{ulem}
\usepackage{makecell}
\usepackage{graphicx} 
\usepackage{xcolor}
\usepackage{tikz}
\usepackage{stfloats}
\usepackage{cuted}
\usepackage{capt-of}
\usepackage{CJKutf8}

\DeclareRobustCommand{\syscomp}[1]{%
\tikz[baseline=(X.base)]\node[
    fill=black!8,
    rounded corners=2.5pt,
    inner xsep=3.6pt,
    inner ysep=1.8pt,
    font=\normalsize \itshape
](X){#1};%
}

\DeclareRobustCommand{\syscompInCaption}[1]{%
\tikz[baseline=(X.base)]\node[
    fill=black!8,
    rounded corners=2.0pt,
    inner xsep=3.5pt,
    inner ysep=1.5pt,
    font=\footnotesize \itshape
](X){#1};%
}

\usepackage{tikz}
\usepackage{xcolor}

\definecolor{viewTagPurple}{HTML}{220B5E}
\definecolor{viewTagBlue}{HTML}{1B4B6E}
\definecolor{viewTagCream}{HTML}{FFE9C2}

\DeclareRobustCommand{\viewtag}[1]{%
  \raisebox{0.05ex}{%
    \tikz[baseline=(char.base)]{
     
      \shade[
        rounded corners=0.35mm,
        left color=viewTagPurple,
        right color=viewTagBlue,
        shading angle=-45
      ]
      (-1.9mm,-1.9mm) rectangle (1.9mm,1.9mm);

      \node[
        text=viewTagCream,
        font=\rmfamily\bfseries\fontsize{9pt}{9pt}\selectfont,
        inner sep=0pt
      ] (char) at (0,-0.03mm) {#1};
    }%
  }%
}

\DeclareRobustCommand{\viewtagInCaption}[1]{%
  \raisebox{-0.08ex}{%
    \scalebox{0.75}{\viewtag{#1}}%
  }%
}

\DeclareRobustCommand{\compotag}[1]{%
  \raisebox{0ex}{%
    \tikz[baseline=(char.base)]{
      \shade[
        left color=viewTagBlue,
        right color=viewTagPurple,
        shading angle=45
      ] (0,0) circle[radius=1.9mm];

      \node[
        text=viewTagCream,
        font=\sffamily\bfseries\fontsize{9pt}{9pt}\selectfont,
        inner sep=0pt
      ] (char) at (0,-0.05mm) {#1};
    }%
  }%
}
\DeclareRobustCommand{\compotagInCaption}[1]{%
  \raisebox{-0.08ex}{%
    \scalebox{0.75}{\compotag{#1}}%
  }%
}

\DeclareRobustCommand{\matchedlineicon}{%
  \raisebox{-0.2ex}{\includegraphics[width=0.9em]{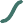}}\nobreak\hspace{0.2em}%
}

\DeclareRobustCommand{\missinglineicon}{%
  \raisebox{-0.2ex}{\includegraphics[width=0.9em]{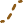}}\nobreak\hspace{0.2em}%
}

\DeclareRobustCommand{\extralineicon}{%
  \raisebox{-0.2ex}{\includegraphics[width=0.9em]{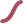}}\nobreak\hspace{0.2em}%
}

\newcommand{\toolname}{\textbf{\textit{GraphQAG}}\xspace}

\begin{document}

\title{\toolname{}: A Knowledge-Graph-Guided Visual Analytics Framework for Question-Answer Pairs Generation}



\author{%
Yize Li,
Ruiqi Yu,
Tianya Pan,
Ningxin Li, 
Songyue Li,
Xiangyang Wu, 
Jinchang Li and Zhiguang~Zhou\textsuperscript{*}

\thanks{Yize Li and Ruiqi Yu contributed equally to this work.}
\thanks{Yize Li, Ruiqi Yu, Tianya Pan, Ningxin Li, Songyue Li, Xiangyang Wu, 
and Zhiguang Zhou are with Hangzhou Dianzi University, Hangzhou, China. E-mail: \{liyize, richyu, pan, lnx, songsongli, wuxy, zhgzhou\}@hdu.edu.cn.}
\thanks{Jinchang Li is with Zhejiang University of Finance \& Economics, Hangzhou, China. E-mail: ljc@zufe.edu.cn.}
\thanks{\textsuperscript{*}Corresponding author: Zhiguang Zhou.}%
}

\markboth{IEEE Transactions on Visualization and Computer Graphics}%
{Li \MakeLowercase{\textit{et al.}}: \toolname{}}


\maketitle

\begin{abstract}
Question--answer (QA) pairs are widely used in knowledge base construction, question--answering systems, and the post-training of large language models (LLMs). However, important knowledge in long documents is often distributed across multiple paragraphs and connected through complex entity relationships. Such fragmented and relational knowledge poses substantial challenges for existing QA generation methods, which often fail to adequately cover core document content, cross-paragraph semantic connections, and multi-entity relationships. We present \toolname{}, a knowledge graph-guided visual analytics framework for generating high-quality QA pairs from long documents. \toolname{} follows a three-stage workflow. First, it constructs a document knowledge graph by segmenting the document into paragraphs and extracting salient entities and relations. Second, it builds a graph-based generation space from entities, relations, and multi-hop paths to constrain and guide LLM-based QA generation. Third, it uses the knowledge graph as an interactive visual representation, enabling users to explore document knowledge structures, inspect the coverage and evidence provenance of generated QA pairs, and iteratively refine the QA pair set through graph-based interactions. We evaluated \toolname{} through a user study with 16 participants, two case studies, and expert interviews. The results indicate that \toolname{} effectively supports users in identifying knowledge coverage gaps, examining generated QA pairs, and refining the QA pair set. These findings demonstrate the usefulness of combining knowledge graphs, LLM-based generation, and visual analytics for producing more comprehensive and trustworthy QA pairs from long documents.

\end{abstract}


\section{Introduction}

\begin{figure*}[t]
    \centering
    \includegraphics[width=\textwidth]{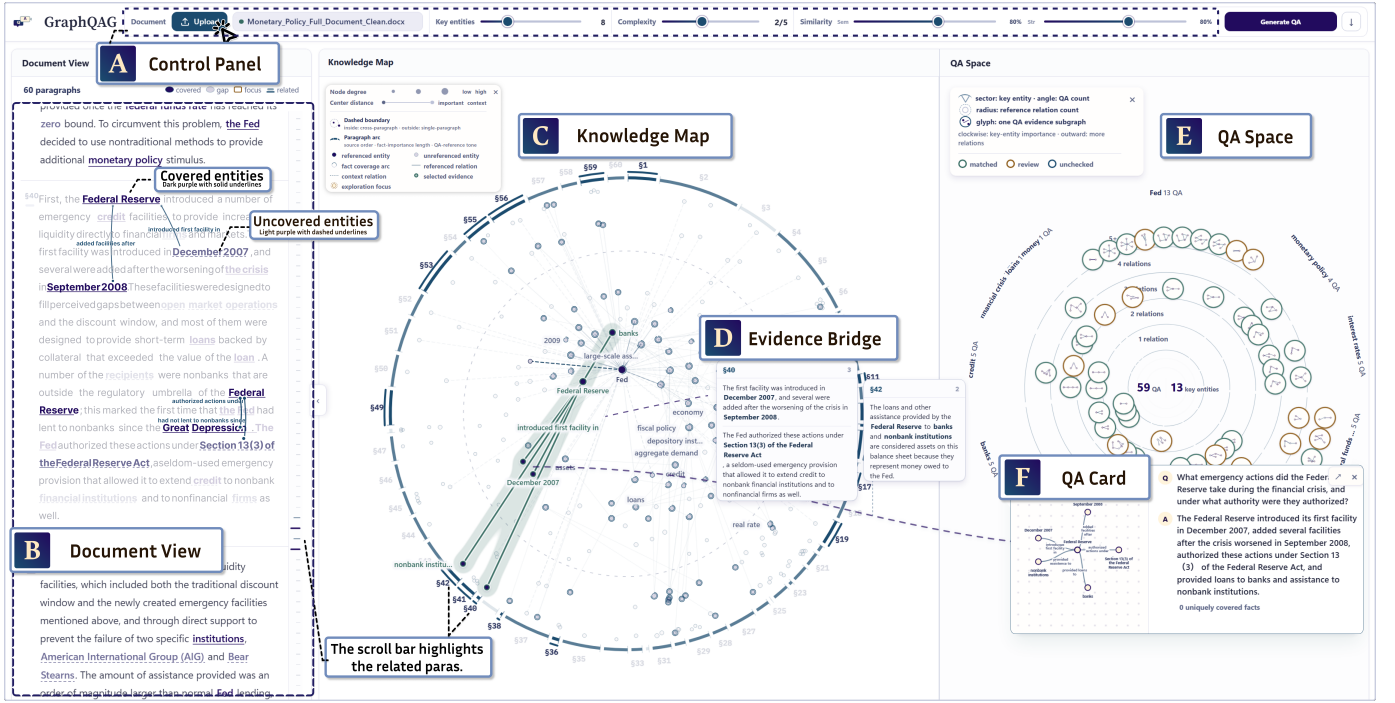}
    \caption{The interface of \toolname{}.
\viewtagInCaption{A} The \syscompInCaption{Control Panel} allows users to upload documents and adjust the system parameters for QAG.
\viewtagInCaption{B} The \textit{Document View} presents the source text, where dark-blue entities indicate knowledge covered by the current QA pair set, while light-blue underlined entities indicate uncovered knowledge.
\viewtagInCaption{C} The \syscompInCaption{Knowledge Map} visualizes the document KG and its current QA coverage.
\viewtagInCaption{D} The \syscompInCaption{Evidence Bridge} links QA pairs to their supporting passages and graph elements.
\viewtagInCaption{E} The \syscompInCaption{QA Space} presents the distribution of generated QA pairs according to key entities and reasoning complexity levels.
\viewtagInCaption{F} The \textit{QA Card} presents the question, answer, and corresponding subgraph of a selected QA pair.}
    \label{fig:systemoverview}
\end{figure*}

\IEEEPARstart{Q}{uestion-answer} (QA) pairs provide a direct and flexible format for organizing document knowledge. They have been widely used in knowledge base construction~\cite{zhong2023kgconstruction}, information retrieval~\cite{karpukhin-etal-2020-dense}, and QA systems~\cite{rajpurkar-etal-2016-squad}. More recently, QA pairs have become an increasingly important source of supervision for the post-training of large language models (LLMs). For example, Phi-4-reasoning is fine-tuned on QA-style prompt--response pairs with teacher-generated reasoning, while DeepSeek-V4 uses domain-specific supervision and teacher outputs to train specialized experts and distill their capabilities into a unified model~\cite{abdin2025phi4reasoning,deepseekai2026deepseekv4highlyefficientmilliontoken}. These developments highlight the importance of question--answer pairs generation (QAG) for constructing high-quality training data and supporting a wide range of downstream applications.

Traditional QAG methods generally follow pipeline-based or end-to-end paradigms~\cite{guo2024survey}. Pipeline-based methods identify candidate answers and generate questions~\cite{duan-etal-2017-question}, whereas end-to-end methods directly generate questions from input contexts, reducing reliance on hand-crafted pipelines~\cite{du-etal-2017-learning,guo2024survey}. Despite their progress, both paradigms often rely on local contexts and surface patterns. As a result, they are less effective when QAG requires distant evidence, multi-entity reasoning, or complex semantic understanding~\cite{liu2024lost}. Without a global representation of document knowledge, existing methods may overlook informative QA pairs, leading to incomplete knowledge coverage and reducing downstream value.

Knowledge graphs (KGs) represent document knowledge as interconnected entities and relations, making cross-passage semantic connections explicit~\cite{ji2022survey}. Such structured representations help models capture long-range dependencies and complex relations beyond local contexts. Recent QAG studies have introduced KGs to guide question generation, using graph structures to organize distributed knowledge and support complex reasoning~\cite{jin2024graphcot,chen2024subgraph}. However, most existing work focuses on model design and generation performance, offering limited support for visually exploring graph-based generation spaces, diagnosing quality issues, and refining QA pair sets. To the best of our knowledge, KG-guided tools that systematically support long-document QAG, assessment, and refinement remain limited.

Based on a literature review and formative interviews with domain experts, we identified three main challenges (\textbf{C1}--\textbf{C3}) for KG-guided visual analytics in long-document QAG. These challenges are as follows: \textbf{(C1)~Translating high-quality QAG goals into a graph-based generation space is challenging.} High-level generation goals do not naturally correspond to concrete graph structures, especially when important knowledge is distributed across paragraphs and connected through multiple relations. \textbf{(C2)~Transforming graph structures into natural-language QA pairs remains challenging.} Graph structures encode relational knowledge, but they do not directly determine question intent, answer boundaries, or linguistic expression. \textbf{(C3)~Assessing and refining generated QA pairs is difficult.} The causes of quality issues are hard to locate without connecting QA pairs to their knowledge structures, making it difficult for users to decide where and how to revise the results.

To address these challenges, we present \toolname{}, a KG-guided visual analytics framework for QAG (Fig.~\ref{fig:systemoverview}). \toolname{} constructs a document knowledge graph and uses its entities, relations, and evidence passages to define a graph-based generation space. This space guides LLM-based QAG and links the generated results back to the underlying graph structures and document evidence. Through coordinated visual views, \toolname{} enables users to explore document knowledge, inspect generated QA pairs, and iteratively refine the QA pair set. We evaluated \toolname{} through a user study with 16 participants, two case studies, and expert interviews. The results show that \toolname{} effectively supports KG-guided QA pairs generation, inspection, and refinement for long documents. The main contributions of this work are as follows: 
\begin{itemize}
\item Through a literature review and collaboration with four domain experts, we identify four core quality objectives for long-document QAG and propose a unified framework that organizes the QAG workflow into generation, assessment, and refinement stages.

\item We design and implement \toolname{}, a KG-guided visual analytics system that operationalizes this framework through document KG construction, graph-based generation space exploration, LLM-based QAG, and evidence-grounded visual inspection.

\item We evaluate \toolname{} through a user study with 16 participants, two case studies, and expert interviews. The results demonstrate its effectiveness and usability in supporting long-document QAG workflows.
\end{itemize}

\section{Related Work}
\label{sec:related_work}

This work is related to prior research on \textit{question--answer pair generation}, \textit{QA pair quality assessment}, and \textit{visual analytics for KG-guided text generation}.

\subsection{Question--Answer Pair Generation}
\label{sec:qag_related_work}

QAG plays an important role in constructing QA systems and creating training data. Existing QAG methods have been broadly developed along pipeline-based, end-to-end, and LLM-assisted paradigms~\cite{mulla2023automatic}. Early studies commonly adopted modular pipelines that decomposed QAG into answer selection, question generation, and quality filtering. For example, Ruan et al.~\cite{ruan2019quizbot} developed QuizBot, which combines predefined QA pairs, semantic similarity, and spaced repetition to support adaptive practice. Lee et al.~\cite{lee2023liquid} proposed LIQUID, which constructs QA datasets from unlabeled text by identifying candidate answer sets and generating corresponding questions. With advances in sequence-to-sequence models, QAG research gradually shifted toward unified end-to-end generation. Ushio et al.~\cite{ushio2023empirical} empirically compared language-model-based QAG methods and showed that unified end-to-end models provide strong and efficient baselines. More recent studies further introduce explicit structures, automated feedback, and LLMs to improve generation quality. Pham et al.~\cite{pham2024graph} used procedural graphs to improve semantic coverage, Wang et al.~\cite{wang2025sea3} combined adversarial generation and automated feedback for long-context QAG, and Yuen et al.~\cite{yuen2025automatic} explored LLM-based synthetic QAG.

In comparison to prior QAG methods that mainly optimize the automatic generation process, our work focuses on supporting users in constructing and refining QA pair sets from long documents. Existing methods can improve generation efficiency and linguistic quality, but they often provide limited support for exploring document-level knowledge structures, controlling the generation scope, and revising generated QA pairs based on uncovered knowledge. \toolname{} addresses this limitation by introducing a KG-guided visual workflow that connects document knowledge structures with QA pairs generation and refinement.

\subsection{Quality Assessment of QA Pairs}
\label{sec:qa_pair_quality_assessment}

QA pair assessment can be considered at two complementary levels: individual-level and collection-level. At the individual level, traditional metrics such as exact match, F1, BLEU, and ROUGE compare generated content with reference text~\cite{rajpurkar-etal-2016-squad,oh2023evaluation}. Although efficient and reproducible, these metrics may penalize valid paraphrases and cannot fully assess answerability or document grounding. Recent studies therefore introduce more targeted evaluation methods, including RQUGE for answerability~\cite{mohammadshahi2023rquge}, QGEval for fluency, relevance, and consistency~\cite{fu2024qgeval}, and RAGAS, RAGChecker, and MiniCheck for document-grounded relevance, faithfulness, and claim support~\cite{es2024ragas,ru2024ragchecker,tang2024minicheck}.

At the collection level, assessment focuses on whether the generated QA pair set provides broad knowledge coverage, sufficient diversity, and an appropriate range of reasoning difficulty. Eo et al.~\cite{eo2023diverse} and Yadav et al.~\cite{yadav-etal-2024-explicit} improve diversity by controlling question forms, answer types, and other generation conditions. Tatarinov et al.~\cite{tatarinov2025kgqagen} introduce KG-QAGen, which uses knowledge-graph structures to generate questions involving different multi-hop paths, set operations, and answer sizes. Amanlou et al.~\cite{amanlou2026knight} propose KNIGHT, combining KG-guided generation with adaptive hardness calibration to control the difficulty of multiple-choice questions.

Existing research on QA pair assessment mainly provides automatic scores or predefined generation controls. In contrast, \toolname{} supports interactive assessment and refinement of QA pair sets. It helps users examine individual QA pairs together with their supporting evidence and graph structures, while also revealing collection-level patterns such as uncovered knowledge and potential redundancy. This design enables users to move from passive quality assessment to targeted revision of the generated QA pair set.

\subsection{Visual Analytics for KG-Guided Text Generation}
\label{sec:visual_analytics_kg_generation}

Visual analytics has been used to make KG-guided generation more visible and inspectable. Existing systems mainly support three types of workflows. First, some systems expose the structured knowledge used before or during generation. LinkQ presents LLM-generated KG queries through editable query and graph views before summarizing retrieved facts~\cite{li2024linkq}. NetMe~2.0 lets users select entities and paths from a document-derived biomedical KG and generates cited explanations from associated evidence~\cite{dimaria2024netme}. Second, some systems help users inspect how generated content is grounded in KG facts. KNowNet maps triples extracted from generated responses to validated KG facts in a progressive graph~\cite{yan2025knownet}, while HealthGenie combines retrieved subgraph views with preference controls for refining KG-grounded recommendations~\cite{gao2025healthgenie}. Third, visualization has also been used to inspect reasoning and retrieval--generation pipelines. KRAGEN visualizes KG-supported reasoning as a graph of thoughts~\cite{matsumoto2024kragen}, and VisPile links generated summaries and questions to KG facts and source passages~\cite{coscia2026vispile}. RAGViz and RAGTrace expose how retrieved contexts influence generation~\cite{wang2024ragviz,cheng2025ragtrace}, while XGraphRAG and RAGExplorer combine graph views, inference traces, matrices, and Sankey diagrams to diagnose retrieval pipelines and failures~\cite{wang2025xgraphrag,tian2026ragexplorer}.

Unlike these systems, \toolname{} focuses on KG-guided QAG from long documents. Existing visual analytics systems mainly support inspecting individual generated responses, reasoning traces, or retrieval processes, but provide limited support for coordinating multiple QA pairs as a document-grounded collection. Our work addresses this gap by integrating document KG construction, graph-based generation-space exploration, and inspection and refinement at both the individual and collection levels into one visual analytics workflow.

\section{Formative Study}
\label{sec:formative_study}

Through literature review and interviews with 4 experts, we identified QA pair quality dimensions and requirements.

\subsection{Collaboration with Domain Experts}
\label{sec:expert_collaboration}

We worked with four domain experts to understand current workflows, common quality problems, and tool needs in long-document QAG. We first reviewed prior studies on QA pair quality and used the results as an initial framework for discussion. We then conducted several rounds of semi-structured interviews. The interviews focused on the experts' practical experience, quality criteria, and expectations for supporting tools. By comparing findings from the literature with expert feedback, we identified the main quality goals and design requirements for \toolname.

\subsubsection{Literature Review}
\label{sec:formative_literature_review}

We conducted a literature review using both search-driven and reference-driven methods. The search employed keywords such as ``question generation,'' ``question--answer pair generation,'' ``QA pair evaluation,'' ``knowledge graph question generation,'' and ``visual analytics for text generation.'' We searched major digital libraries and scholarly databases, including ACL Anthology, IEEE Xplore, ACM Digital Library, and Google Scholar. We focused on studies related to three topics: automatic QAG, QA pair quality assessment, and KG-guided or visualization-supported text generation. Papers were included if they proposed QAG methods, discussed quality metrics or evaluation criteria for QA pairs, used knowledge structures to guide generation, or provided visual/interactive support for inspecting generated text. After the initial search, we further examined references in the identified papers to include relevant studies. This process yielded 51 representative papers. Based on these studies, we summarized six candidate quality dimensions for document QA pairs, including \textit{Knowledge Coverage},\textit{ Reasoning Complexity},\textit{ Evidence Grounding} , \textit{Question--Answer Alignment}, \textit{Diversity}, and\textit{ Non-redundancy,} as shown in Table~\ref{tab:quality_dimensions}.

\begin{table*}[t]
\caption{Quality dimensions of document QA pairs summarized from prior studies.}
\label{tab:quality_dimensions}
\centering
\footnotesize
\renewcommand{\arraystretch}{1.15}
\begin{tabular}{
>{\raggedright\arraybackslash}m{0.19\textwidth}
>{\raggedright\arraybackslash}m{0.61\textwidth}
>{\centering\arraybackslash}m{0.12\textwidth}
}
\hline
\textbf{Quality Dimension} & \textbf{Description} & \textbf{References} \\
\hline
\textit{\textbf{Knowledge Coverage}}
& A QA pair set should cover important knowledge units and relationships in the source document rather than concentrate on only a small number of salient entities or passages.
& \cite{pham2024graph,lee2023liquid} \\

\textit{\textbf{Reasoning Complexity}}
& Each QA pair should involve a meaningful reasoning structure over relations and evidence, such as relation understanding, cross-paragraph synthesis, or multi-hop reasoning.
& \cite{yang-etal-2018-hotpotqa,trivedi-etal-2022-musique,kumar2019difficulty} \\

\textit{\textbf{Evidence Grounding}}
& Each QA pair should be supported by identifiable source passages, with its answer and reasoning process traceable to the corresponding evidence.
& \cite{yang-etal-2018-hotpotqa,es2024ragas,tang2024minicheck} \\

\textit{Question--Answer Alignment}
& Each QA pair should maintain semantic alignment between its question and answer, such that the answer directly addresses the question and satisfies its semantic constraints.
& \cite{fu2024qgeval,mohammadshahi2023rquge,es2024ragas} \\

\textit{Diversity}
& A QA pair set should include varied question forms, answer types, knowledge perspectives, and reasoning patterns.
& \cite{eo2023diverse,yadav-etal-2024-explicit} \\

\textit{\textbf{Non-redundancy}}
& Each QA pair should be structurally and semantically distinct from other pairs, avoiding substantial overlap in graph structures, supporting evidence, or question--answer content.
& \cite{yadav-etal-2024-explicit,pham2024graph,sultan-etal-2020-importance} \\
\hline
\end{tabular}
\end{table*}

\subsubsection{Iterative Interviews}
\label{sec:iterative_interviews}
Building on the literature-derived quality dimensions, we conducted iterative interviews to examine their applicability to long-document QAG practices and to elicit concrete requirements for a supporting tool.

\textbf{Participants.}
We invited 4 experts (\textbf{E1}--\textbf{E4}, aged 30--50) to participate in the semi-structured interviews. The experts have complementary backgrounds in natural language processing, visual analytics, data engineering, and industrial applications.
\textbf{E1} and \textbf{E2} are university professors. \textbf{E1} specializes in natural language processing (10 years exp.) and has extensive experience in document understanding, question generation, and LLM applications. \textbf{E2} specializes in visualization and visual analytics (12 years exp.) and has research experience in KG visualization and human--AI collaborative analysis.
\textbf{E3} and \textbf{E4} work in industry. \textbf{E3} is an AI data engineer at an internet company and has experience in question answering systems and domain-model fine-tuning (8 years exp.). \textbf{E4} is a senior engineer in an enterprise digital transformation department (10 years exp.) and has experience in enterprise knowledge base construction and business document management.

\textbf{Procedure.}
We collected expert feedback through iterative semi-structured interviews, with each round organized around 3 topics: 
(1) We asked the experts to describe their current QAG workflows based on their research or professional experience. The discussion covered document selection, prompt design, result inspection, and the revision of unsatisfactory QA pairs.
(2) We discussed common quality problems in existing QAG methods. We presented the candidate dimensions identified in the literature review. We then asked the experts whether these dimensions were suitable for long-document QAG and what problems they often observed in generated QA pairs.
(3) We asked the experts to describe the functions they expected from a supporting tool. We recorded their feedback on generation workflows, quality criteria, and system functions. After each round, we summarized the comments and used them to refine the questions for the next round. This process helped confirm the problems and design directions.

\subsection{Findings}
\label{sec:formative_findings}

The formative study showed that long-document QAG is not a simple one-step generation task. Although existing methods can quickly produce QA pairs, the generated results often require further inspection, verification, and refinement before they can be used as high-quality QA data. Based on the expert feedback, we summarized 3 main findings:

\textbf{(1) Experts emphasized quality control in long-document QAG.}
The experts agreed that one-step generation cannot guarantee usable QA pairs and that users need explicit support for inspection and refinement. After reviewing the six candidate dimensions in Table~\ref{tab:quality_dimensions}, they prioritized four quality goals: \underline{\textbf{\textit{Knowledge Coverage}}}, \underline{\textbf{\textit{Reasoning Complexity}}}, \underline{\textbf{\textit{Evidence Grounding}}}, and \underline{\textbf{\textit{Non-redundancy}}}. They did not prioritize \textbf{\textit{Question--Answer Alignment}} because current LLMs generally produce answers semantically aligned with their questions, making severe alignment errors uncommon. They also did not prioritize \textbf{\textit{Diversity}}, because varied question forms and answer types do not necessarily increase the knowledge value of the QA pair set. These dimensions guided the system requirements for generation-space construction, evidence-grounded generation, and multi-level assessment.

\textbf{(2) Experts viewed KGs as a bridge between document understanding and QAG.}
The experts noted that important knowledge in long documents is often distributed across paragraphs and connected through entities, relations, and evidence snippets. A document KG can make such distributed knowledge explicit and help users understand the document as a connected knowledge space. It can also provide a structured scope for generation, allowing QA pairs to be generated from selected graph structures rather than isolated text fragments. Therefore, the experts considered KG guidance useful for supporting QA pairs that better reflect the document's core knowledge and relational structure.

\textbf{(3) Experts considered iterative assessment and refinement necessary.}
The experts emphasized that high-quality QAG usually requires multiple rounds of checking and revision. Generated QA pairs may omit important knowledge, rely on insufficient evidence, involve shallow reasoning, or repeat similar content. These problems are difficult to address through automatic generation alone. Users therefore need to inspect generated QA pairs, trace their supporting knowledge, remove unsuitable results, and supplement missing content. This finding motivated us to design \toolname{} as an iterative workflow in which users can assess and refine QA pairs through graph-based visual feedback.

\subsection{Design Requirements}
\label{sec:design_requirements}

The experts (\textbf{E1}--\textbf{E4}) agreed that KG-guided QAG should progress from document understanding and generation-scope specification to evidence-grounded generation and iterative refinement. Based on their feedback, we derived six requirements across three dimensions: \textbf{constructing a graph-based generation space}, \textbf{grounding QA generation in evidence}, and \textbf{assessing and refining QA pairs at multiple levels}.

\textbf{D1. Constructing a graph-based generation space.}

\textbf{R1. Construct a document KG with source-evidence links.}
\textbf{E1} noted that important knowledge in long documents is often scattered across paragraphs and must be connected through entities and relations. \textbf{E4} emphasized evidence traceability: \textit{``If a QA pair cannot be traced back to the source document, it is difficult to verify and reuse it.''} Thus, the system should construct a document KG that links graph facts to supporting source passages, enabling users to trace graph structures back to the original evidence.

\textbf{R2. Map generation goals to an inspectable graph-structured space.}
\textbf{E1} explained that QA pairs may rely on different structures, including local facts, multi-hop paths, and synthesis relations. \textbf{E2} stressed that these structures should be visible and controllable: \textit{``Users should see which entities, relations, and paths are being used before the system generates QA pairs.''} Thus, the system should organize generation goals as graph structures for inspection, selection, and adjustment.

\textbf{D2. Grounding QA generation in evidence.}

\textbf{R3. Guide QAG with graph structures and source evidence.}
\textbf{E1} observed that an implicit generation scope may cause LLMs to omit important content or combine unrelated information. \textbf{E3} noted that QA pairs for model training should be generated within explicit knowledge and evidence boundaries. Accordingly, the system should use selected entities, relations, paths, and source passages as generation constraints.

\textbf{R4. Expose the evidence and reasoning structure of each QA pair.}
\textbf{E1} emphasized that multi-relation QA pairs should expose intermediate reasoning structures. \textbf{E2} explained, \textit{``A question and an answer alone do not show why the result is reliable. Users need to see the graph structure and supporting passages.''} The system should therefore connect each QA pair to its source evidence, related graph elements, and reasoning path.

\textbf{D3. Assessing and refining QA pairs at multiple levels.}

\textbf{R5. Assess quality at both individual and set levels.}
\textbf{E1} emphasized inspecting individual QA pairs for evidence grounding and reasoning completeness. \textbf{E2} suggested using visual summaries to assess the overall QA pair set from a global perspective, while \textbf{E4} focused on whether important business knowledge was covered without unnecessary repetition. Thus, the system should support individual-level evidence and reasoning inspection, together with set-level diagnosis of knowledge coverage and redundancy.

\textbf{R6. Support targeted graph-based refinement.}
\textbf{E2} suggested linking user feedback to graph structures to make refinement goals visible. \textbf{E3} noted, \textit{``If only a few QA pairs have problems, regenerating the batch is inefficient and gives users little control.''} \textbf{E4} expected users to generate additional pairs for missing knowledge. The system should therefore support targeted addition, revision, repair, and removal based on graph structures, source evidence, and diagnostic results.

\begin{figure*}[t]
    \centering
    \includegraphics[width=\textwidth]{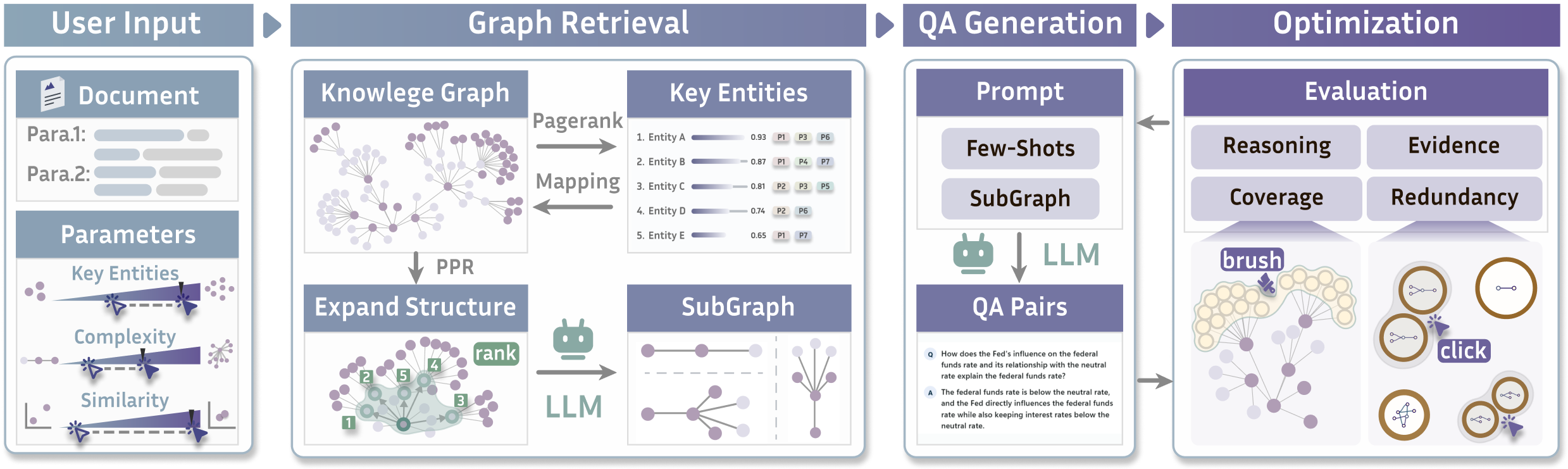}
\caption{The pipeline of \toolname{}. Given a source document and user-specified generation parameters, the system constructs a document KG, identifies and ranks key entities, expands their local graph structures to retrieve generation subgraphs, and uses them to guide LLM-based QAG. It then assesses the generated QA pairs along four quality dimensions and supports iterative refinement through visual interaction.}
    \label{fig:pipeline}
\end{figure*}

\section{Method}

This section details the methodology. Fig.~\ref{fig:pipeline} shows the framework building a document KG, deriving generation units, and verifying and refining QA pairs through graph and evidence feedback.

\subsection{Key Entity Identification}
\label{sec:key_entity_identification}
The system first divides the text into several semantically coherent segments. It then uses GraphRAG \cite{edge2024graphrag} to extract entities and their relationships from each segment, merges semantically equivalent entities across different segments, and constructs a KG covering the entire document (\textbf{R1}).

KGs typically contain a large number of entities, making it difficult to identify valuable target entities. To address this issue, \toolname{} first systematically evaluates entity importance. Specifically, it uses PageRank~\cite{brin1998anatomy} with paragraph priors to measure the importance of entities in the KG (\textbf{R2}). Let $\mathcal{P}$ denote the nonempty set of paragraphs in the document that contain entities, and let $V_p$ denote the set of distinct entities contained in paragraph $p$. The system first assigns each paragraph an equal initial probability mass of $1/|\mathcal{P}|$ and then distributes this mass uniformly among the entities in that paragraph. Accordingly, the paragraph prior probability of entity $v_i$ is defined as follows:
\begin{equation}
\pi(v_i)
=
\sum_{\substack{p\in\mathcal{P}\\v_i\in V_p}}
\frac{1}{|\mathcal{P}|\cdot |V_p|}
\label{eq:eq1}
\end{equation}
Its PageRank score at each iteration is defined as follows:

\begin{equation}
PR_i^{(t+1)}
=
\bigl(1-d+dD^{(t)}\bigr)\pi_i
+
d\sum_{j\in\mathcal N_i}
\frac{PR_j^{(t)}}{\deg(v_j)}
\label{eq:eq2}
\end{equation}
where $PR^{(t)}$ denotes the importance score of an entity at the $t$-th iteration, and $d$ is the damping factor. $\mathcal{N}(v_i)$ denotes the set of distinct entities adjacent to entity $v_i$ in the KG, and $\deg(v_j)$ denotes the number of entities adjacent to entity $v_j$. $D^{(t)}$ denotes the total probability mass of all dangling entities at the $t$-th iteration. The probability mass of dangling entities is redistributed according to the paragraph prior $\pi$. The algorithm is considered to have converged when the difference between two consecutive iterations falls below a predefined threshold. Finally, \toolname{} ranks the entities in descending order based on their final importance scores and selects the top $k$ entities to form the set of key entities.

\subsection{Local Subgraph Retrieval}
\label{sec:local_subgraph_retrieval}

The set of key entities represents the core knowledge in a KG, but the scope of knowledge it delineates is implicit. Therefore, after identifying the key entities, the system further retrieves entities with high structural relevance to each key entity and constructs local subgraphs for QAG (\textbf{R3}). We employ Personalized PageRank (PPR) \cite{jeh2003scaling} to measure the structural relevance between entities in the KG and the key entities. The system treats each key entity as a seed and performs personalized random walks over the complete KG. PPR scores are computed iteratively, and convergence is reached when the change between two consecutive iterations falls below a predefined threshold. For each key entity, the system ranks all entities in descending order according to their PPR scores and prompts the user to specify $k$, the number of top-ranked entities to retrieve. The selected top-$k$ entities, together with the key entity, form a candidate node set. The user also selects a complexity level from 1 to 5, which determines the maximum length of a relation chain originating from the key entity. Higher levels include more distant entities and support questions requiring longer reasoning chains. Lee et al. \cite{lee-etal-2025-grade} observe that QA accuracy generally decreases as reasoning depth increases from 1 to 5 hops. Therefore, to limit error accumulation while retaining support for complex multi-hop questions, we cap the relation-chain length at five hops. Finally, the system filters the candidate nodes according to the selected complexity level and retains their factual relations from the original KG, thereby constructing a local subgraph centered on the key entity.

\subsection{Subgraph-Guided QAG}
\label{sec:subgraph_guided_qag}
To enable subgraph-guided QAG, the system completes two preparatory steps: \textit{reference subgraph} construction and question specification definition (\textbf{R4}). It encodes entities, relations, and their indices as structured inputs and uses an LLM to select, from the relation set, edges that can jointly support the same question. The objective of this construction process is to extract necessary and interconnected facts while preserving the structural information required for reasoning. After obtaining the candidate \textit{reference subgraphs}, the system validates the relation indices, removes duplicate edges, and retains the set of edges connected to the key entity. The degree of overlap between different \textit{reference subgraphs} is measured using the Jaccard similarity of their edge sets. When the similarity exceeds a predefined threshold, only the subgraph containing more information is retained. The \textit{reference subgraph} construction and deduplication procedure is summarized in Algorithm~\ref{alg:evidence-subgraph}.

\begin{algorithm}[htb]
\caption{Reference Subgraph Construction}
\label{alg:evidence-subgraph}
\begin{algorithmic}[1]
\Require Relation set $\mathcal{R}$, candidate index sets $\mathcal{I}$,
central entity $e_c$, threshold $\tau$
\Ensure Deduplicated reference subgraphs $\mathcal{G}$

\State $\mathcal{G} \gets \emptyset$

\ForAll{$I \in \mathcal{I}$}
    \State $E' \gets \operatorname{Unique}
    \bigl(\{\mathcal{R}[i] \mid i \in I \land 1 \leq i \leq |\mathcal{R}|\}\bigr)$
    \State $E \gets
    \{(u,r,v)\in E' \mid u,v\in
    \operatorname{CC}(e_c;E')\}$
    \State $G \gets
    \bigl(\{u \mid (u,r,v)\in E\}\cup
    \{v \mid (u,r,v)\in E\},E\bigr)$

    \If{$E = \emptyset$}
        \State \textbf{continue}
    \EndIf

    \State $(s,G^\ast) \gets
    \max\limits_{G_j=(V_j,E_j)\in\mathcal{G}}
    \left(
    \dfrac{|E\cap E_j|}{|E\cup E_j|},G_j
    \right)$

    \If{$\mathcal{G}=\emptyset$ \textbf{or} $s\leq\tau$}
        \State $\mathcal{G}\gets\mathcal{G}\cup\{G\}$
    \ElsIf{$|E|>|E^\ast|$}
        \State $\mathcal{G}\gets
        (\mathcal{G}\setminus\{G^\ast\})\cup\{G\}$
    \EndIf
\EndFor

\State \Return $\mathcal{G}$
\end{algorithmic}
\end{algorithm}

On this basis, the system converts each \textit{reference subgraph} into a corresponding explicit question specification that defines the factual scope covered by the question, the reasoning relations to be expressed, and the answer slot. Each question specification includes the central entity, candidate answers, evidence relations, fact identifiers, and source paragraphs. During the QAG stage, the LLM converts the question specification into QA pairs. Its inputs include the question specification, the \textit{reference subgraph}, and the relevant source sentences. The QAG prompt template constrains the question objective, answer slot, and scope of evidence use, and the resulting QA pair is returned in a structured format.

\subsection{Reverse Verification and Optimization}
\label{sec:reverse_verification}
To evaluate the quality of the result, the system reconstructs a subgraph using only the QA pairs and the KG, and validates it by comparison with the \textit{reference subgraph} (\textbf{R5}). It segments the QA pairs into semantically complete clauses and performs entity linking based on semantic similarity. It then integrates textual similarity with graph structure to match and rank candidate facts corresponding to each clause. The system prioritizes high-confidence facts and connects facts that share the same entities. For entities that remain disconnected but are explicitly mentioned in the QA pairs, the system traverses the shortest paths to establish auxiliary connections and recover the implicit reasoning process. Finally, it selects a high-confidence connected subgraph as the reconstruction result.

\begin{figure}[hbt]
    \centering
    \includegraphics[width=0.85 \linewidth]{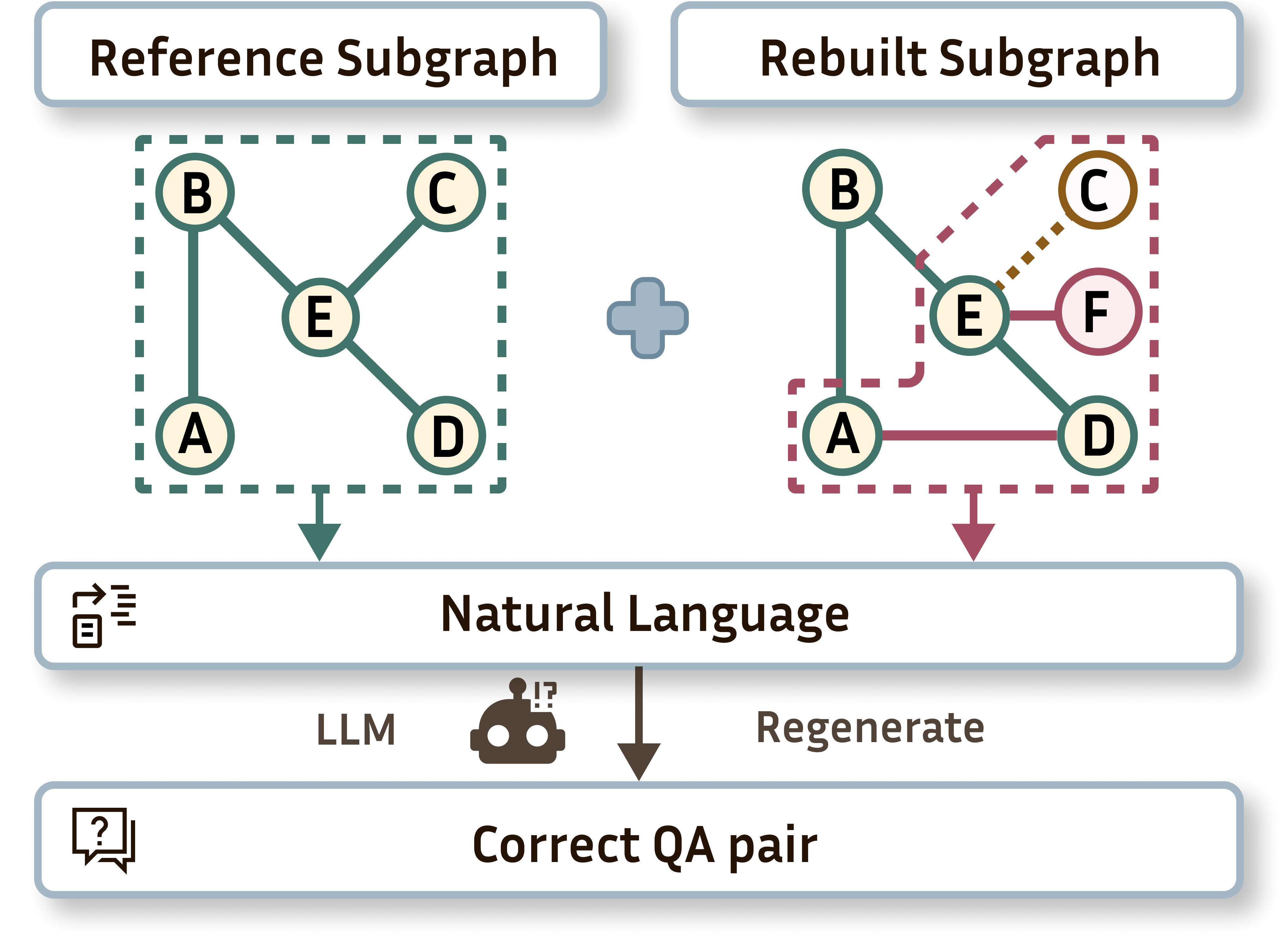}
    \captionsetup{skip=-2pt}
    \caption{Comparison and repair of QA subgraphs. The \textit{ reference subgraph} (left) and the \textit{rebuilt subgraph} from a generated pair (right) represent entities as nodes and semantic relations as links. Their comparison reveals a missing relation between E and C, an extraneous entity F, and altered relations involving A, B, and D. These discrepancies are verbalized and provided to an LLM to guide QA pair regeneration.}
    \label{fig:method}
\end{figure}

The system compares the rebuilt subgraph with the \textit{reference subgraph} by the fact set, entity set, source passages, and graph structure, and classifies the differences into six categories: exact match, fact omission, fact addition, fact substitution, irrelevant evidence, and indeterminate. A QA pair is valid only if the facts in the pair can occur in the KG and the reconstructed result is consistent with the reference evidence in both factual content and structure.

For each QA pair, the system jointly inputs the \textit{reference subgraph}, the rebuilt subgraph, and the original QA pair into LLM for correction (Fig.~\ref{fig:method}). Based on the differences between the two subgraphs in entities, relations, and structure, the model removes unsupported facts, adds missing information, and corrects erroneous relations. 
Each revised QA pair is then subjected to reverse validation again, and is included as valid samples only when its rebuilt subgraph is consistent with the reference evidence (\textbf{R6}).

\section{\toolname{}}

This section presents \toolname{}. We first give a system overview, then describe its visual components, interactions, and implementation.

\begin{figure*}[t]
    \centering
    \includegraphics[width=\textwidth]{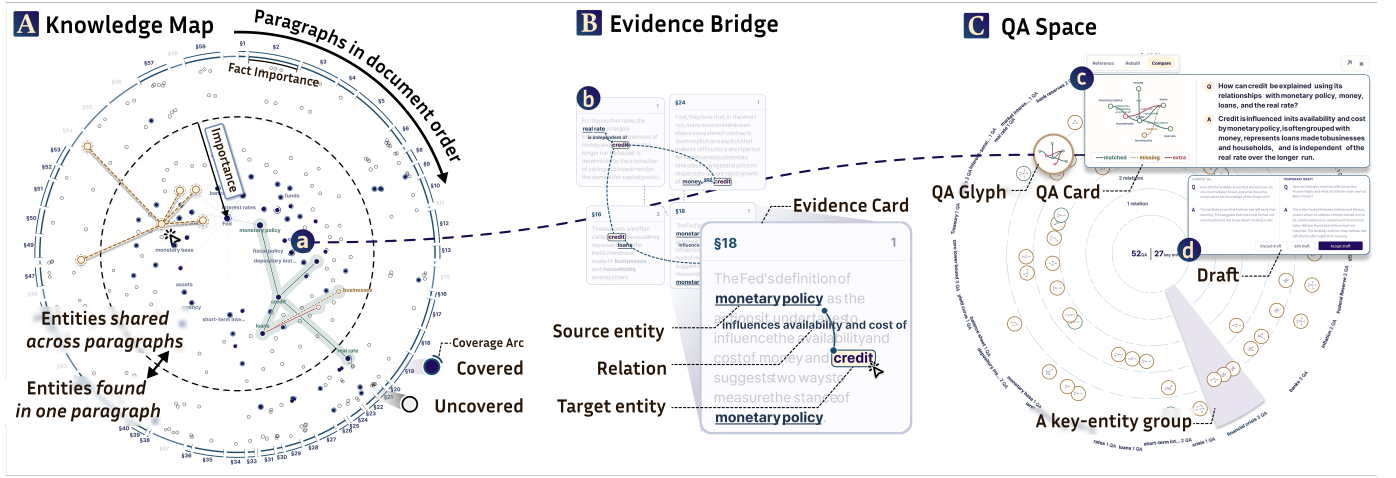}
   \caption{\viewtagInCaption{A} The \syscompInCaption{Knowledge Map} organizes entities by source paragraph and highlights a selected \textit{reference subgraph}, and users can explore the reasoning structure on it~\compotagInCaption{a}. \viewtagInCaption{B} The \syscompInCaption{Evidence Bridge} presents supporting source passages in linked~\compotagInCaption{b} \syscompInCaption{Evidence Cards}. \viewtagInCaption{C} The \syscompInCaption{QA Space} organizes QA pairs across three semantic zoom levels. A selected QA pair expands into a~\compotagInCaption{c} \textit{QA Card}. \compotagInCaption{d} supports the configuration and inspection of an LLM-assisted repair draft before acceptance. The forward flow from \compotagInCaption{a} to \compotagInCaption{c} supports QA generation. The reverse flow supports evidence-based verification.}
\label{fig:visualdesign}
\end{figure*}

\subsection{System Overview}
\label{sec:system_overview}

We present \toolname{}, a visual analytics system for generating, inspecting, and refining a QA pair set from long documents. The interface provides a \textit{Control Panel} (Fig.~\ref{fig:systemoverview}-\viewtag{A}) and a \textit{Document View} (Fig.~\ref{fig:systemoverview}-\viewtag{B}). The visual analysis uses 3 coordinated components: the \syscomp{Knowledge Map} (Fig.~\ref{fig:systemoverview}-\viewtag{C}), \syscomp{Evidence Bridge} (Fig.~\ref{fig:systemoverview}-\viewtag{D}), and \syscomp{QA Space} (Fig.~\ref{fig:systemoverview}-\viewtag{E}).

The \textit{Control Panel} supports document upload, \textit{key-entity settings}, \textit{complexity settings}, \textit{similarity thresholds}, and \textit{QA generation}. The \textit{key-entity setting} determines how many PageRank-ranked entities are used as retrieval seeds, following the method in Section~\ref{sec:key_entity_identification}. The \textit{complexity setting} controls the hop range for local subgraph retrieval described in Section~\ref{sec:local_subgraph_retrieval}. The \textit{similarity thresholds} support the identification of potentially redundant QA pairs. The \textit{QA generation} control runs the subgraph-guided QAG method described in Section~\ref{sec:subgraph_guided_qag}.
The \textit{Document View} presents the source text and indicates covered and uncovered entities (\textbf{R1}). The \syscomp{Knowledge Map} presents the document KG, key entities, reference subgraphs, and local QA-reference coverage. It makes the graph-based generation targets constructed in Section~\ref{sec:subgraph_guided_qag} inspectable (\textbf{R2}, \textbf{R3}). The \syscomp{Evidence Bridge} links selected graph facts and QA pairs to their supporting source passages (\textbf{R4}). The \syscomp{QA Space} organizes the QA pair set at the \textit{Distribution Level}, \textit{Redundancy Level}, and \textit{Detail Level} (Fig.~\ref{fig:qaspace}). These levels present QA distribution, potential redundancy, and the individual validation results produced by the reverse verification method in Section~\ref{sec:reverse_verification} (\textbf{R5}).

\subsection{Visual Design}
\label{sec:visual_design}

The interface uses three coordinated visual components. The \syscomp{Knowledge Map} presents document knowledge and its source locations. The \syscomp{QA Space} organizes generation goals and QA pairs. The \syscomp{Evidence Bridge} connects the selected structure to its source text. The components share paragraph, entity, relation, generation-target, and QA identifiers. A selection in one component is reflected in the others. This design follows the guidance for coordinated multiple views~\cite{baldonado2000guidelines}.

\subsubsection{Knowledge Map}

Users need to see how knowledge from different paragraphs is connected and where each relation comes from (\textbf{R1}). The \syscomp{Knowledge Map} uses a radial node-link layout. 
The outer ring follows the order of the source paragraphs, with each sector representing one paragraph.
Its angular width encodes the total importance weight of the graph facts linked to that paragraph. 
We divide entities into two types based on the source paragraphs in which they appear 
. The \syscomp{Knowledge Map} places these two types in separate regions. \textit{Shared entities} appear in two or more paragraphs and link information across paragraphs. They are placed in the inner region. \textit{Local entities} appear in only one paragraph and are placed in the outer region within their corresponding paragraph sectors. This design makes cross-paragraph connections visible while keeping local entities close to their source paragraphs (Fig.~\ref{fig:visualdesign}-\viewtag{A}).
The paragraph sectors follow the source order around the outer ring, which preserves the document sequence~\cite{wattenberg2002arc}. The separate inner and outer regions reduce visual overlap between shared and local entities~\cite{shneiderman2006network}. Within each region, the distance from the center encodes the entity importance score defined in Eq.~\ref{eq:eq2}, following prior radial graph layouts~\cite{brandes2011radial}. Entities with higher scores are placed closer to the center.
For an entity \(v\), let \(I_G(v)\) be the set of unique generation-eligible facts incident to \(v\) in the document KG. Let \(E_R\) be the set of unique facts included in the r\textit{eference subgraphs} of the current QA pair set. Let \(w(f)\) denote the normalized importance weight of fact \(f\). We define the local QA-reference coverage of \(v\) as
\begin{equation}
C_{\mathrm{QA}}(v)
=
\frac{
\sum_{f \in I_G(v) \cap E_R} w(f)
}{
\sum_{f \in I_G(v)} w(f)
}.
\label{eq:local-qa-fact-coverage}
\end{equation}
Each canonical fact is counted once. The score ranges from 0 to 1 and is shown as a percentage. The length of the arc around each entity encodes this score~\cite{skau2016arcs}. A full ring indicates \(100\%\) coverage, while no arc indicates \(0\%\) coverage. A short arc reveals a gap in the reference scope and supports targeted refinement (\textbf{R5}, \textbf{R6}).
Links represent graph relations. 
Their width and opacity reflect fact importance, coverage, and whether the relation belongs to the current generation space. 
These visual channels help users distinguish knowledge importance, coverage, and generation status~\cite{mackinlay1986automating}. 

\subsubsection{Evidence Bridge}

Users need to trace the graph facts associated with a QA pair to their source text while keeping the selected QA pair and subgraph visible (\textbf{R1}, \textbf{R4}). The \syscomp{Evidence Bridge} (Fig.~\ref{fig:visualdesign}-\viewtag{B}) draws a dashed curve from the selected subgraph in the \syscomp{Knowledge Map} to its QA pair in the \syscomp{QA Space}. It creates one \syscomp{Evidence Card} (Fig.~\ref{fig:visualdesign}-\compotag{b}) for each source paragraph linked to the selected graph facts. Evidence excerpts from the same paragraph appear in the same card. The header shows the paragraph number, the section title when available, and the number of linked graph facts.

The cards are arranged in compact rows around the middle of the curve. A thin dashed line connects each card to the curve. This arrangement shows evidence from several paragraphs together with the selected QA pair and subgraph~\cite{steinberger2011context}. Entity mentions use bold blue text. When users focus on an entity mention, the system dims unrelated text. 
Within the same card, labeled blue arcs connect the focused entity to mentions of directly related entities. 
If the focused entity appears in more than one card, dashed curves connect its mentions across the cards. 
These curves indicate that the same entity appears in different paragraphs. 
This design helps users follow an entity across graph facts and source passages~\cite{stasko2008jigsaw}.

These visual links provide the source context for inspecting the corresponding QA pair in the \syscomp{QA Space}.

\subsubsection{QA Space}

\begin{figure*}[t]
    \centering
    \includegraphics[width=\textwidth]{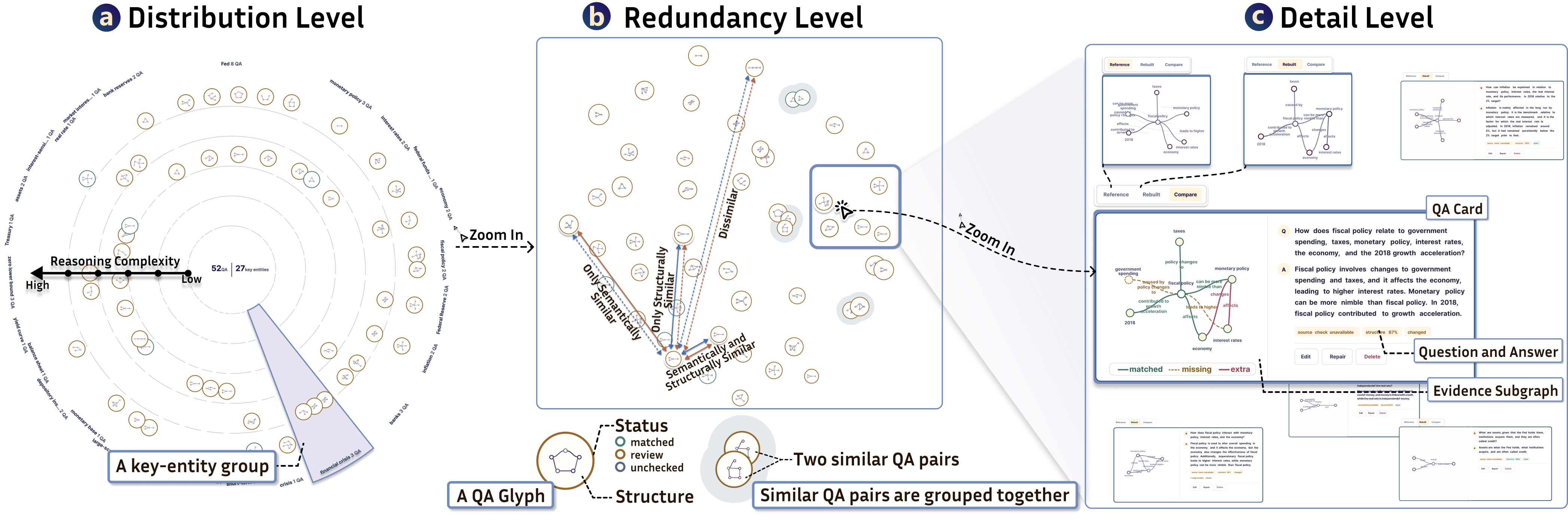}
    \caption{The \syscompInCaption{QA Space} supports progressive inspection of a QA pair set at three semantic zoom levels. \compotagInCaption{a} The \textit{Distribution Level} groups \textit{QA glyphs} by key entity. Sector angle encodes QA count. Radial distance encodes the number of relations in the \textit{reference subgraph}. Outline color indicates validation status. \compotagInCaption{b} The \textit{Redundancy Level} places \textit{QA glyphs} by semantic and structural similarity. Pale blue envelopes mark potentially redundant groups. \compotagInCaption{c} The \textit{Detail Level} expands a selected glyph into a \textit{QA Card}. The card shows the question, answer, validation summary, and the \textit{Reference}, \textit{Rebuilt}, and \textit{Compare} views.}
    \label{fig:qaspace}
\end{figure*}

The \syscomp{QA Space} organizes the QA pair set from overview to detail while preserving the set context (\textbf{R5}). It provides three levels of semantic zoom: the \textit{Distribution Level}, the \textit{Redundancy Level}, and the \textit{Detail Level} (Fig.~\ref{fig:qaspace}). This three-level design follows established guidance on overview, semantic zoom, and details on demand~\cite{shneiderman1996eyes,bederson1996pad}.

At the \textit{Distribution Level}, each sector groups \textit{QA glyphs} by one key entity (Fig.~\ref{fig:qaspace}-\compotag{a}). Sector angle encodes the number of QA pairs in the group. Radial distance encodes reasoning-structure size, measured by the number of relations in the \textit{reference subgraph}. Glyph outlines indicate whether a QA pair is matched, needs review, or is unchecked. These encodings show how QA pairs are distributed across key entities and how their reasoning-structure size and validation status differ.

At the \textit{Redundancy Level}, each \textit{QA glyph} represents one QA pair (Fig.~\ref{fig:qaspace}-\compotag{b}). \textit{QA glyphs} are positioned through a projection of semantic and structural similarity. Two \textit{QA glyphs} are close only when both their QA content and \textit{reference subgraph} structures are similar. A low score on either measure keeps them apart. The horizontal and vertical directions have no independent meaning. The orange and blue guide lines in Fig.~\ref{fig:qaspace}-\compotag{b} illustrate semantic and structural similarity, respectively. Solid and dashed lines indicate whether the corresponding similarity exceeds its threshold. The four combinations distinguish pairs that are similar in one measure, both measures, or neither. Pale blue bubbles surround QA pairs that exceed both similarity thresholds and may be redundant.

At the \textit{Detail Level}, a selected \textit{QA glyph} expands into a \textit{QA Card} as users zoom in (Fig.~\ref{fig:qaspace}-\compotag{c}). 
At the \textit{Detail Level}, a selected \textit{QA glyph} expands into a \textit{QA Card} as users zoom in (Fig.~\ref{fig:qaspace}-\compotag{c}). 
The card shows the question, answer, subgraph views, and validation summaries. 
The \textit{Reference} view shows the \textit{reference subgraph} used for generation. 
The \textit{Rebuilt} view shows the \textit{subgraph reconstructed} from the QA pair by reverse verification in Section~\ref{sec:reverse_verification}.
The \textit{Compare} view overlays the reference and rebuilt subgraphs in the same layout (\textbf{R4}). 
\matchedlineicon Green solid lines show matched facts. 
\missinglineicon Amber dashed lines show missing facts. 
\extralineicon Red solid lines show extra facts. 
Node outlines use green, amber, and red for the same three states. Missing nodes have dashed outlines. 
This layout makes structural differences easy to compare~\cite{gleicher2011visual}. 
When unsupported content is detected, the card reports the number of ungrounded clauses (\textbf{R5}).

\subsection{Interactions}
\label{sec:interactions}
Interactions connect diagnosis with refinement across the source text, graph structure, QA pairs, and the QA pair set.

\textbf{Cross-View Exploration.}
Hovering over an entity, relation, paragraph, or QA pair provides a temporary preview. Clicking keeps the related focus active. Selecting a graph element reveals its source context (\textbf{R1}). Selecting a QA pair highlights its graph structure, displays its \syscomp{Evidence Cards}, and locates its evidence in the \textit{Document View} (\textbf{R4}).

\textbf{Semantic Zoom.}
Users scroll in the \syscomp{QA Space} to move between the \textit{Distribution Level}, \textit{Redundancy Level}, and \textit{Detail Level} (\textbf{R5}). Hovering over a \textit{QA glyph} shows a preview. Clicking keeps the preview visible. Double-clicking opens its full \textit{QA Card} at the \textit{Detail Level}. The \syscomp{Knowledge Map} also supports panning and zooming.

\textbf{Targeted QA Generation.}
Users choose \textit{Generate QA }to create an initial QA pair set. They can select text in the \textit{Document View} and choose \textit{Generate Here}. The system selects the generation target that best overlaps the entities in that text. In the \syscomp{Knowledge Map}, users can build a reasoning trail by selecting highlighted relations. Each selection adds one fact and updates recommendations. The system sends the confirmed \textit{reference subgraph} and minimal source evidence to the LLM (\textbf{R3}). The generated QA pair is validated and added to the \syscomp{QA Space} with validation status.

\textbf{Individual QA Inspection and Refinement.}
Users open a \textit{QA Card} and switch among the \textit{Reference}, \textit{Rebuilt}, and \textit{Compare} views (\textbf{R4}). Hovering over a graph element highlights its counterpart in the \syscomp{Knowledge Map}. During Repair, users can select diagnostic clauses and inspect matched facts and source evidence. Editing creates a temporary draft. For Repair, the LLM regenerates a temporary draft from the fixed \textit{reference subgraph}, minimal source evidence, and selected \textit{repair} constraints. The system reconstructs and validates each draft. Users accept a valid draft or discard it (\textbf{R6}). The current QA pair remains unchanged until acceptance. Delete removes the QA pair from the current set.

\textbf{Set-Level Assessment and Completion.}
Users inspect local QA fact coverage in the \syscomp{Knowledge Map}. At the \textit{Distribution Level}, they inspect QA distribution, reasoning complexity, and validation status. At the \textit{Redundancy Level}, glyph proximity and pale blue envelopes reveal potentially redundant QA pairs (\textbf{R5}). Users can adjust the similarity thresholds, QAG for uncovered facts, and edit, repair, or delete problematic pairs (\textbf{R6}). The system updates the visual encodings after each accepted change.

\subsection{Implementation}

The \toolname{} frontend is built with Vue~3 and TypeScript, and Pinia manages shared state. D3~\cite{bostock2011d3} lays out the \syscomp{Knowledge Map}, and Vue Flow presents the \syscomp{QA Space}. The backend uses FastAPI to provide document processing and QA services. GraphRAG~\cite{edge2024graphrag} extracts entities and relations from documents. Sentence-Transformers~\cite{reimers2020multilingual,song2020mpnet} links QA text to graph elements. DeepSeek-V4-Flash~\cite{deepseekai2026deepseekv4highlyefficientmilliontoken} supports QA generation, verification, and repair.

\section{Evaluation}
We evaluated \toolname{} from two research questions:

\textbf{RQ1.} How does \toolname{} improve the effectiveness and usability of long-document QAG?

\textbf{RQ2.} How does \toolname{} help users produce higher-quality QA pair sets?

To answer these questions, we conducted three complementary evaluations: 
(1) A within-subjects user study comparing \toolname{} with a Baseline system in terms of task effectiveness, usability, and QA pair quality (\textbf{RQ1} and \textbf{RQ2}); 
(2) Two case studies showing how users used \toolname{} to discover knowledge gaps, examine evidence, and iteratively refine QA pair sets in realistic scenarios (\textbf{RQ2}); 
and 
(3) Expert interviews gathering feedback on the system design and the practical value of KG-guided QAG (\textbf{RQ1} and \textbf{RQ2}).

\begin{figure*}[t]
    \centering
    \includegraphics[width=\textwidth]{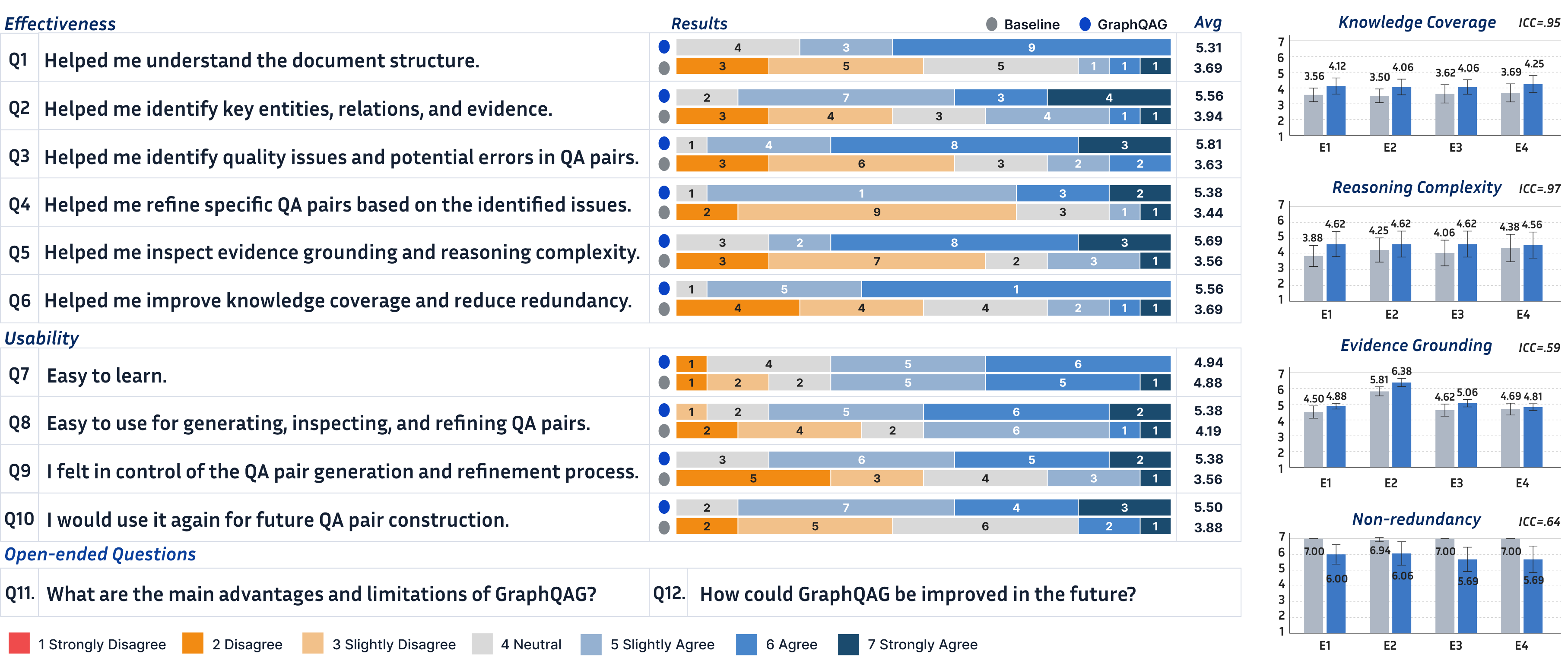}
    \caption{User study results. The left panel summarizes participants' responses on Effectiveness (\textbf{Q1}--\textbf{Q6}) and Usability (\textbf{Q7}--\textbf{Q10}). Paired stacked bars show the seven-point Likert-scale distributions for \toolname{} and the baseline, and the Avg column reports the corresponding mean scores; \textbf{Q11}--\textbf{Q12} present the two open-ended questions. The right panels compare expert assessments of the final QA pair sets under the two conditions across four quality dimensions. \textbf{E1}--\textbf{E4} denote the four experts; bars show mean ratings across the 16 participant outputs, and error bars indicate 95\% confidence intervals.
}
    \label{fig:like}
\end{figure*}

\subsection{User Study}
\label{sec:user_study}

To answer \textbf{RQ1} and \textbf{RQ2}, we conducted a within-subjects user study comparing \toolname{} with Baseline.
We examined participants' subjective experience and the quality of their final QA pair sets.

\textbf{Participants and Task.}
We recruited 16 participants (\textbf{P1}--\textbf{P16}, 8 female and 8 male). All participants had prior experience using LLMs but had not previously used \toolname{}. Among them, 6 participants (\textbf{P5}--\textbf{P7}, \textbf{P9}, and \textbf{P14}--\textbf{P15}) reported relatively high familiarity with KGs, with ratings of 4 or 5 on the five-point familiarity scale. The experiment employed a within-subjects design in which each participant completed two QAG tasks, one using \toolname{} and the other using the Baseline condition. Two long documents from different domains were selected as the experimental materials, including Arthur Conan Doyle's detective story \textit{A Scandal in Bohemia}~\cite{doyle_scandal_bohemia}  and a monetary policy report from the GovReport dataset~\cite{huang-etal-2021-efficient}. The two documents were approximately comparable in length, number of paragraphs, number of entities and relations, knowledge density, and the extent of cross-paragraph reasoning required. Under each condition, participants used one document to generate, review, and refine a set of QA pairs. To ensure a fair comparison, \toolname{} and the Baseline used the same underlying LLM, DeepSeek-V4-Flash~\cite{deepseekai2026deepseekv4highlyefficientmilliontoken}, and participants were required to keep the key generation parameters consistent across the two conditions, including the number of key entities and the complexity hop setting. The Baseline was implemented as an LLM-plus-prompt interaction condition, in which participants used prompts to guide the same underlying LLM in generating QA pairs from the source document. Participants could read the document, write prompts, request QA pair generation, and manually revise or select the generated results. However, the Baseline did not provide \toolname{}'s graph-structured generation space, graph-to-evidence mapping, multi-level quality assessment, or interactive refinement capabilities. A Latin square design~\cite{salinas2024latin} was adopted to counterbalance the order of experimental conditions, the order of document presentation, and the assignment of documents to conditions (4 groups $\times$ 4 participants).

\textbf{Procedure.}
Each participant completed \toolname{} and Baseline in sequence. The study lasted approximately 2 hours. Before each condition, participants received a brief tutorial, practiced with a separate text, and spent 15 minutes reviewing the task document and its topic and structure. They then used the assigned tool to generate, verify, and refine QA pairs before submitting a final QA pair set. Both conditions used the same task requirements and time limits.
After each condition, participants completed the same 10-item questionnaire on a seven-point Likert scale, covering Effectiveness (\textbf{Q1}--\textbf{Q6}) and Usability (\textbf{Q7}--\textbf{Q10}). We averaged \textbf{Q1}--\textbf{Q6} and \textbf{Q7}--\textbf{Q10} for each participant and condition to obtain the two scores. After the \toolname{} condition, participants answered two open-ended questions (\textbf{Q11}--\textbf{Q12}) about its advantages, limitations, and potential improvements.

The same four experts from the formative study (\textbf{E1}--\textbf{E4}) independently evaluated 32 final QA pair sets. Before rating, they spent approximately 30 minutes reviewing the two source documents, identifying key knowledge points, and resolving differences. They created a shared checklist for each document, covering key entities, relations, supporting evidence, and expected reasoning links. We anonymized and randomized all sets without revealing their conditions. Experts rated each set on a seven-point scale for knowledge coverage, reasoning complexity, evidence grounding, and non-redundancy. For each dimension, we assessed inter-rater reliability using a two-way random-effects, absolute-agreement, average-measures intraclass correlation coefficient, ICC(2,4), with 95\% confidence intervals~\cite{kumar2024icc,feng2021interval}. We averaged four expert ratings for each set and dimension to obtain expert quality scores.

\textbf{Quantitative results.}
\toolname{} received significantly higher ratings than the Baseline on both Effectiveness and Usability (Fig.~\ref{fig:like}).
For Effectiveness, \toolname{} achieved a median score of 5.67 ($IQR=0.50$), compared with 3.42 ($IQR=1.21$) for the Baseline. A two-sided Wilcoxon signed-rank test showed a significant difference ($W=3.00$, $p<.001$, $r_{rb}=.96$).
For Usability, median scores were 5.25 ($IQR=0.63$) for \toolname{} and 4.00 ($IQR=1.19$) for the Baseline, also showing a significant difference ($W=2.00$, $p=.004$, $r_{rb}=.95$).
At the item level, the largest descriptive mean differences were found in identifying quality issues and potential errors (\textbf{Q3}: 5.81 vs. 3.63), inspecting evidence grounding and reasoning complexity (\textbf{Q5}: 5.69 vs. 3.56), and refining QA pairs (\textbf{Q4}: 5.38 vs. 3.44).
In contrast, the two conditions received similar ratings for learnability (\textbf{Q7}: 4.94 vs. 4.88).
These results show that \toolname{} primarily improved participants' ability to inspect and refine QA pairs, while maintaining learnability comparable to the Baseline.

In expert assessment (Fig.~\ref{fig:like}), inter-rater agreement was excellent for knowledge coverage ($\mathrm{ICC}(2,4)=.95$, 95\% CI $[.91, .97]$) and reasoning complexity ($\mathrm{ICC}(2,4)=.97$, 95\% CI $[.95, .98]$), but moderate for evidence grounding ($\mathrm{ICC}(2,4)=.59$, 95\% CI $[.12, .81]$) and non-redundancy ($\mathrm{ICC}(2,4)=.64$, 95\% CI $[.37, .81]$). The ratings for evidence grounding and non-redundancy showed greater between-expert variability than knowledge coverage and reasoning complexity.

Based on the ratings averaged across the four experts (Fig.~\ref{fig:like}), \toolname{} received significantly higher ratings than the Baseline for reasoning complexity (\toolname{}: $Mdn=4.25$; Baseline: $Mdn=4.00$; $W=0.00$; $p_{\mathrm{adj}}=.015$; $r_{rb}=1.00$) and evidence grounding (\toolname{}: $Mdn=5.25$; Baseline: $Mdn=4.88$; $W=17.00$; $p_{\mathrm{adj}}=.049$; $r_{rb}=.68$). \toolname{} showed a higher median knowledge-coverage rating than the Baseline, but the difference did not remain statistically significant after Holm correction. (\toolname{}: $Mdn=4.50$; Baseline: $Mdn=3.50$; $W=36.00$; $p_{\mathrm{adj}}=.097$; $r_{rb}=.47$). In contrast, the Baseline received a significantly higher rating for non-redundancy (\toolname{}: $Mdn=5.75$; Baseline: $Mdn=7.00$; $W=1.50$; $p_{\mathrm{adj}}=.003$; $r_{rb}=-.98$). 

The quantitative results show that \toolname{} improved participants' perceived effectiveness and the quality of the generated QA pair sets. To better understand these results, we further analyzed participants' comments and interaction behaviors, and identified two recurring themes.

\textbf{Graph Structure as a Scaffold for QA Generation.} Participants reported that \toolname{} organized entities, relations, paths, and supporting evidence into an inspectable graph structure, making the scope of generation more explicit. Compared with the Baseline, where generation requirements were primarily expressed through natural-language prompts, \toolname{} externalized these requirements as visible and manipulable graph elements. Participants could inspect the document-level knowledge structure, identify relevant entities and relations, and determine which knowledge should be included before initiating or refining generation. \textbf{P11} noted, \textit{“The graph structure captures document relations better than plain-text methods.”} This finding helps explain \toolname{}'s higher ratings for knowledge understanding, generation scope control, and evidence-grounded QAG.

\textbf{Multi-Level Views Support Verification and Refinement.} The coordinated views linked QA pairs, their graph structures, and supporting source passages. They enabled participants to trace answers to evidence, assess evidence grounding, and inspect associated reasoning structures. At the set level, the quality overview revealed coverage gaps, missing reasoning structures, and potentially redundant pairs. It helped participants decide whether to add, revise, or remove QA pairs. \textbf{P3} observed, \textit{``With the Baseline, I could only review the results one by one, but \toolname{} directly showed me which content had not yet been covered.''} However, some participants needed time to learn the system and requested clearer onboarding, more explicit explanations of system functions, and stronger visual feedback during QA inspection and refinement.

\subsection{Case Study}

This section presents two cases from the user study. Case 1 shows how \textbf{P8} identified coverage gaps, generated QA pairs for uncovered entities, and removed redundancy. Case 2 shows how \textbf{P15} connected evidence across paragraphs and repaired a suspicious QA pair. We analyzed the recorded interactions and generated results for both cases.

\begin{figure*}[t]
    \centering
    \includegraphics[width=\textwidth]{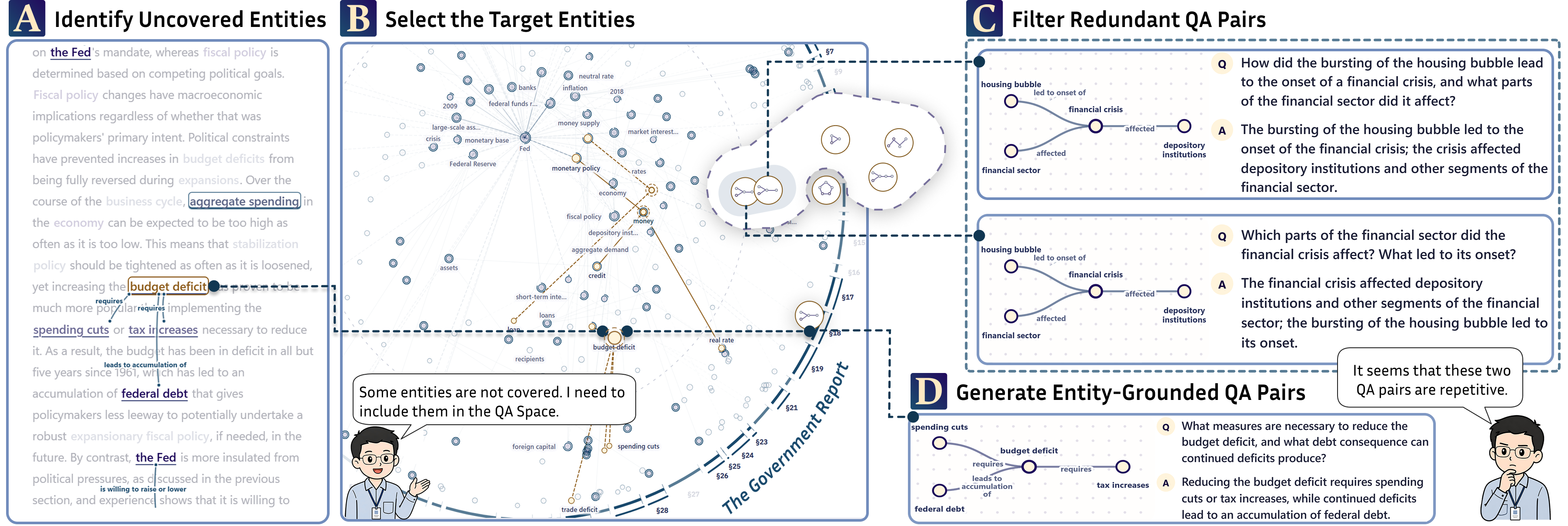}
    \caption{Case 1: Charting knowledge gaps for targeted QA refinement. \viewtagInCaption{A} \textbf{P8} identifies uncovered entities in the \textit{Document View}. \viewtagInCaption{B} The synchronized \syscompInCaption{Knowledge Map} provides an interactive selection mechanism for adding entities that require additional QA coverage. \viewtagInCaption{C} Within the \syscompInCaption{QA Space}, \textbf{P8} identifies overlapping QA pairs through textual and structural comparison. \viewtagInCaption{D} The \syscompInCaption{QA Space} presents the newly generated entity-grounded QA pairs.}
    \label{fig:case1}
\end{figure*}

\subsubsection{Case 1. Charting Knowledge Gaps for Targeted QA Refinement}
\label{sec:case1}

\textbf{P8}, a data annotation engineer with experience in knowledge base construction, was asked to generate a QA pair set from a 60-paragraph monetary policy report in the GovReport dataset. The report described institutions, policy instruments, implementation mechanisms, and economic consequences distributed across multiple sections. \textbf{P8} uploaded the document, set the number of key entities to 9, and selected complexity level 2 on a five-level scale. \toolname{} then constructed the KG and generated 59 candidate QA pairs.

\textbf{Cross-view inspection of the initial QA pair set.}
\textbf{P8} first examined the \syscomp{QA Space}, which organized the 59 candidate pairs by key entities (Fig.~\ref{fig:systemoverview}-\viewtag{E}). Large groups around entities such as \textit{Fed} and \textit{monetary base} suggested that the central monetary-policy concepts had received substantial coverage. Rather than treating the number of generated pairs as evidence of completeness, \textbf{P8} compared their distribution with the highlighted source content in the \textit{Document View} (Fig.~\ref{fig:systemoverview}-\viewtag{B}) and the corresponding entities in the \syscomp{Knowledge Map} (Fig.~\ref{fig:systemoverview}-\viewtag{C}). In the \textit{Document View}, \textbf{P8} observed that many entities were shown in light blue, indicating that they appeared in the document but had not yet been covered by any generated QA pair. The same pattern was also visible in the \syscomp{Knowledge Map}, where many nodes remained light blue. This cross-view comparison helped \textbf{P8} recognize that some document knowledge was still uncovered. As \textbf{P8} observed, \textit{``These questions seem to cover core concepts such as the Fed and the monetary base, but some entities are not covered. I need to include them in the QA Space.''}

\textbf{Targeted generation for uncovered entities.}
To address these gaps, \textbf{P8} selected several uncovered entities of interest in the \textit{Document View}(Fig.~\ref{fig:case1}-\viewtag{A}). When \textbf{P8} clicked some entities, the system highlighted the supporting passage. The synchronized \syscomp{Knowledge Map} isolated the corresponding local subgraph containing these entities and relations (Fig.~\ref{fig:case1}-\viewtag{B}). This provided \textbf{P8} with a concrete, evidence-grounded scope for targeted generation. After inspecting the highlighted passage and local graph structure, \textbf{P8} clicked the \textit{Generate QA} button to generate additional QA pairs. The system regenerated 2 QA pairs and added them to the \syscomp{QA Space} (Fig.~\ref{fig:case1}-\viewtag{D}). One of the generated questions was \textit{``What measures are necessary to reduce the budget deficit, and what debt consequence can continued deficits produce?''}, whose answer combined the three relations in the selected subgraph. \textbf{P8} verified the answer against both the highlighted passage and the graph structure.

\textbf{Redundancy removal through subgraph comparison.}
After confirming that the entities he considered important had been covered, \textbf{P8} became concerned that the expanded QA pair set might contain repetitive questions. He therefore revisited the \syscomp{QA Space} and explored the redundancy layer. By inspecting the distribution of redundancy indicators, \textbf{P8} found that five groups of potentially redundant QA pairs were enclosed by bubbles. One of these groups involved two QA pairs about \textit{financial crisis}(Fig.~\ref{fig:case1}-\viewtag{C}). The two QA pairs exhibited highly similar wording and were associated with essentially identical subgraphs. 
The first asked how the bursting of the housing bubble triggered the financial crisis and which components of the financial sector were affected, whereas the second reversed the order of these two inquiries. 
Their answers conveyed the same substantive information: the bursting of the housing bubble precipitated the onset of the financial crisis, which subsequently affected depository institutions as well as other segments of the financial sector. After comparing their texts and graph structures, \textbf{P8} concluded, \textit{``It seems that these two QA pairs are repetitive.''}
Following the system's redundancy cue, he removed one of them and continued refining the QA pair set. The final set was reduced to 57 QA pairs. Although it was smaller than the initial set of 59 candidates, it covered the entities that \textbf{P8} considered important and contained fewer redundant pairs.

\begin{figure*}
    \centering
    \includegraphics[width=0.96\textwidth]{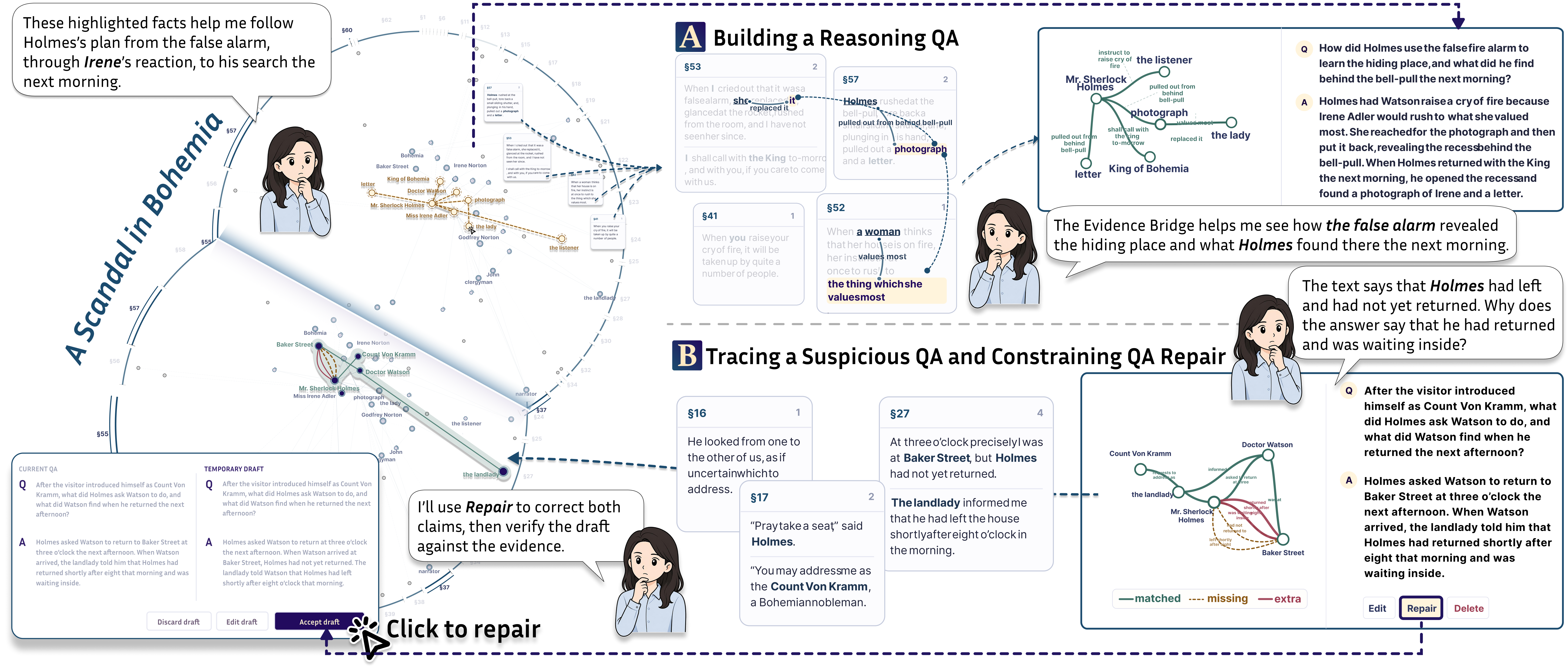}
\caption{Case 2: Constructing and inspecting cross-paragraph QA pairs.
\viewtagInCaption{A} \textbf{P15} explores the \syscompInCaption{Knowledge Map} and selects facts to build a reasoning structure. The selected facts are highlighted in yellow. The \syscompInCaption{Evidence Bridge} brings together their source passages, which \textbf{P15} checks before generating the QA pair.
\viewtagInCaption{B} \textbf{P15} opens a suspicious QA pair and inspects it in the \textit{Compare} mode. Two missing and two extra relations lead \textbf{P15} back to the source passages. \textbf{P15} then uses \textit{Repair} and verifies the temporary draft before accepting it.}
    \label{fig:case2}
\end{figure*}

\subsubsection{Case 2. Constructing and Inspecting Cross-Paragraph QA Pairs}

\textbf{P15}, a Ph.D researcher in narrative visualization, explored the 62 paragraphs of \textit{A Scandal in Bohemia}. \textbf{P15}'s session included two illustrative episodes that capture the generation and inspection workflow of \toolname{}. 
We report how \textbf{P15} constructed a QA pair from distributed clues and how \textbf{P15} inspected and repaired a suspicious result.

\textbf{Building a reasoning QA from distributed clues.} In the \syscomp{Knowledge Map}, \textbf{P15} selected \textit{photograph}, which connected the false fire alarm, Irene Adler's reaction, and Holmes's later search (Fig.~\ref{fig:case2}-\viewtag{A}). The system highlighted nearby candidate facts in yellow. \textbf{P15} checked each candidate in the linked \textit{Evidence Cards} before adding it to the reasoning path. The suggestions updated after each selection. He finally selected six facts about Holmes's plan, Irene's reaction to the alarm, and the later discovery behind the bell-pull. The \syscomp{Evidence Bridge} arranged their supporting passages in source order. Based on this path, \textbf{P15} generated the question \textit{``How did Holmes use the false fire alarm to learn the hiding place, and what did he find behind the bell-pull the next morning?''} The answer explained how the alarm revealed the hiding place and how Holmes later found a photograph and a letter. \textbf{P15} then inspected the pair in the \textit{QA Card}. The \textit{Reference} and \textit{Rebuilt} subgraphs aligned, and the evidence supported the event order. He therefore kept the QA pair without repair.

\textbf{Tracing a suspicious QA and constraining Repair.} After constructing the first QA pair, \textbf{P15} reviewed the existing results in the \syscomp{QA Space}. A yellow outer ring drew \textbf{P15}'s attention to one \textit{QA glyph}. The ring did not mean that the QA pair was incorrect. It indicated that the \textit{Reference} and \textit{Rebuilt} subgraphs were not fully aligned. \textbf{P15} therefore opened the \textit{QA Card} to locate the difference. The QA pair asked about \textit{Holmes's request to Watson and what Watson found at Baker Street the next afternoon}. Its answer claimed that \textit{Holmes had returned shortly after eight and was waiting inside} (Fig.~\ref{fig:case2}-\viewtag{B}). Although the answer read fluently, the \textit{Compare} mode showed four matched relations, two missing relations, and two extra relations. The missing relations stated that Holmes had left shortly after eight and had not returned by three. The extra relations instead stated that Holmes had returned shortly after eight and was waiting inside. Because these structural differences did not alone prove that the answer was wrong, \textbf{P15} checked the linked \textit{Evidence Cards}. The first passage confirmed that the visitor used the name Count Von Kramm. The next passage showed that Holmes asked Watson to return at three o'clock the following afternoon. 
The final passage stated that Holmes had not returned when Watson arrived and that the landlady reported that Holmes had left shortly after eight that morning.
\textbf{P15} therefore found that the answer had reversed Holmes's departure into a return and had added an unsupported claim that Holmes was waiting inside. \textbf{P15} selected \textit{Repair}, and the system used the reference relations and their source passages to produce a temporary draft. The revised answer stated that Holmes asked Watson to return at three, but had not yet returned when Watson arrived. It also retained the landlady's report that Holmes had left shortly after eight. \textbf{P15} checked the draft against the \syscomp{Evidence Cards} and the \textit{Compare} mode. \textbf{P15} added the revised QA pair to the final QA set only after confirming that every statement was supported by the source text.

\subsection{Expert Interview}
\label{sec:expert_interview}
\textbf{Participants and Procedure.}
To complement the findings for \textbf{RQ1} and \textbf{RQ2}, we conducted semi-structured interviews with four experts from the formative study (\textbf{E1}--\textbf{E4}). Each session lasted about 80 minutes and included two phases. First, we introduced \toolname{} and asked the experts to complete a 20-minute hands-on task to familiarize themselves with the workflow and views. Second, we conducted an independent interview based on their hands-on experience. We then presented user-study materials, including action sequences, intermediate and final QA pair sets, questionnaire responses, and four-dimensional quality scores, to support a follow-up discussion on observed user behaviors and quantitative results. We reviewed the transcripts, analyzed the independent interviews and follow-up discussions separately, and synthesized recurring observations into three themes.

\textbf{Understanding the QA Generation Space.}
The experts noted that the \syscomp{Knowledge Map} and \syscomp{QA Space} helped users understand the document and refine the generation scope. As \textbf{E1} explained, \textit{``In prompt-based workflows, users often begin generating questions before understanding the document as a whole. The knowledge graph first provides a map of the available knowledge.''} \textbf{E4} further noted that users could see which entities and relations were included in the generation scope and which content remained uncovered, while suggesting clearer links among the overview, selected subgraph, and QA pair set. When reviewing the action sequences, the experts observed that participants often used the \syscomp{Knowledge Map} to locate uncovered knowledge before adding QA pairs. Overall, making the generation space visible helped users select content more purposefully through explicit knowledge structures rather than natural-language prompts alone.

\textbf{Inspecting Evidence and Reasoning.}
The experts valued the coordinated inspection of QA pairs, source passages, and graph structures. \textbf{E3} remarked, \textit{``When the supporting evidence and the QA subgraph are presented together, missing intermediate relations and unsupported reasoning steps become much easier to identify.''} \textbf{E4} similarly explained, \textit{``In the baseline, users can ask the model to check or revise a QA pair, but they still need to identify the relevant document content and describe the problem through conversation. Here, users can directly inspect the answer, its source evidence, and the corresponding graph structure together.''} The experts viewed this coordinated inspection as a key advantage of \toolname{}, because it makes the basis of each QA pair easier to trace and verify. They also noted that the ``Reference Subgraph'' is particularly valuable for QA pairs that require reasoning across multiple passages or relationships.

\textbf{Supporting Iterative Refinement.}
The experts agreed that the Add QA, Edit, Repair, and Delete operations address different refinement needs during QAG. As \textbf{E2} explained, \textit{``Adding a missing question, revising a weak QA pair, and removing redundant content represent different decision-making processes. The system explicitly distinguishes these refinement objectives.''} Compared with the baseline, these targeted operations reduce reliance on prompt-based problem descriptions and make refinement more direct and controllable. \textbf{E3} also endorsed the human-in-the-loop Repair function: \textit{``By comparing the original and revised QA pairs, users can inspect the changes and decide whether to accept the revision.''} The experts further noted that Knowledge Coverage and Non-redundancy cues help users locate specific deficiencies and choose appropriate refinement actions. Together, these features support a more controlled and efficient refinement process.

\textbf{Overall Assessment.}
Overall, the experts reached broad agreement that \toolname{} supports document understanding, generation scope specification, evidence and reasoning inspection, QA pair set evaluation, and iterative refinement. Compared with the baseline, \toolname{} provides more structured support across these activities and makes the QAG process more transparent and inspectable. Taken together, \toolname{} is particularly suitable for structured and evidence-grounded QAG, where users need to define generation goals, verify supporting evidence, and iteratively refine the QA pair set.

\section{Discussion}

We synthesize findings from the formative study, user study, case studies, and expert interviews, and discuss the implications, limitations, and future directions of our work.

\subsection{General Discussion}

Rather than detailing individual findings, we highlight three cross-cutting themes from user study, case studies, and expert interviews on our framework's support for QA pair generation and refinement in long documents.

\textbf{Graph-Structured Generation Scaffolding.}
Across studies, we found that the intermediate representations introduced in our framework—including key entities, reference subgraph, and rebuilt subgraph—helped users make otherwise implicit generation goals explicit. Instead of relying solely on linear document reading or repeatedly revising prompts, participants inspected the graph overview, selected relevant entities and relations, and traced them back to supporting passages. In case studies, these representations helped participants identify important knowledge regions and construct QA pairs involving information distributed across different paragraphs. These findings suggest that graph structures serve a dual role: they constrain the knowledge used by the LLM and provide users with an editable scaffold for understanding and controlling the QAG process.

\textbf{Quality-Oriented QA Pair Set Construction.}
We further found that participants treated generated QA pairs as a connected set rather than isolated outputs. They examined whether the set covered important entities, relations, and reasoning paths, while checking individual pairs for evidence support and semantic overlap. Participants used uncovered graph regions to guide additional generation and compared graph structures with source passages to identify unsupported or repetitive content. The resulting QA pair sets showed stronger reasoning quality and evidence grounding. Knowledge coverage was also higher, though not statistically significant. Non-redundancy was lower because some similar QA pairs were retained as complementary. These findings suggest that long-document QAG should be viewed as a set-construction task with interdependent and sometimes competing quality objectives, rather than repeated generation of individually acceptable questions.

\textbf{Feedback-Guided Human--LLM Collaboration.}
We observed that our workflow establishes a clear division of labor between users and the LLM. The model translates selected evidence structures and question specifications into candidate QA pairs, while users determine which knowledge is important, whether the selected relations form a meaningful question, and whether the supporting evidence is sufficient. When problems were identified, participants could revise a related entity, path, evidence fragment, or QA pair without regenerating the entire set. This shifted their effort away from prompt-level trial and error and toward higher-level judgments about knowledge value and dataset quality. Our findings suggest that \toolname{} is better understood as a structured authoring and analysis scaffold than as a fully autonomous end-to-end QAG pipeline.

\subsection{Limitations and Future Work}
\label{sec:limitations}

Although the results are encouraging, \toolname{} still has limitations in multimodal document support, expert assessment reliability, and downstream validation. These limitations suggest three directions for future work.

\textbf{Multimodal Document QA Pair Generation.}
\toolname{} currently focuses on text-based long documents and constructs document KGs from paragraphs, entities, relations, and textual evidence spans. However, many real-world documents contain important knowledge in tables, figures, charts, diagrams, and captions, which cannot be fully captured by the current text-only pipeline. As a result, \toolname{} may miss QA pairs that depend on table values, visual trends, figure annotations, or cross-modal references. Future work will incorporate multimodal elements into document KGs, model cross-modal relations between text and visual evidence, and support QA pair generation and inspection grounded in both modalities.

{\textbf{Expert Diversity and Assessment Reliability.}
Our expert assessment involved four experts from natural language processing, visual analytics, data engineering, and industrial applications. Their diverse backgrounds may have contributed to the moderate agreement for evidence grounding and non-redundancy, as they emphasized different aspects of QA quality. For example, similar QA pairs may be viewed as redundant or complementary, and experts may apply different standards for sufficient evidence. Future evaluations should provide clearer rubrics and shared calibration examples while retaining multidisciplinary expertise.}

\textbf{Downstream Validation and Practical Generalizability.}
Our current evaluation mainly examined how \toolname{} supports QA pair construction, assessment, and refinement. We have not yet systematically evaluated whether the generated QA pairs improve downstream applications such as knowledge base construction, domain question answering, or LLM post-training. In addition, the evaluation involved limited documents and domains, which may not fully reflect practical scenarios with heterogeneous structures, specialized terminology, and data-quality issues. Future work will evaluate \toolname{} across broader document types, domains, and downstream tasks, and further examine its impact on factual accuracy, reasoning ability, and domain adaptation.

\section{Conclusion}
We present \toolname{}, a knowledge-graph-guided visual analytics framework for generating high-quality QA pairs from documents. \toolname{} constructs a graph-based generation space composed of entities, relations, structural patterns, and supporting evidence, thereby supporting the generation, inspection, and refinement of QA pairs. The framework consists of three main stages: constructing a document knowledge graph, generating QA pairs based on selected graph structures, and assessing and refining the generated results through visual feedback. We evaluated \toolname{} through a user study with 16 participants, two case studies, and expert interviews. The results show that \toolname{} helps users improve the knowledge coverage, reasoning complexity, evidence grounding, and non-redundancy of generated QA pair sets. By using knowledge graphs as both a grounding structure for generation and an interactive visual medium, this work provides design implications for future research on controllable QA generation and document knowledge construction.

\bibliographystyle{IEEEtran}
\bibliography{bibliography}

\section{Biography Section}

\begin{IEEEbiography}[{\includegraphics[width=1in,height=1.25in,clip,keepaspectratio]{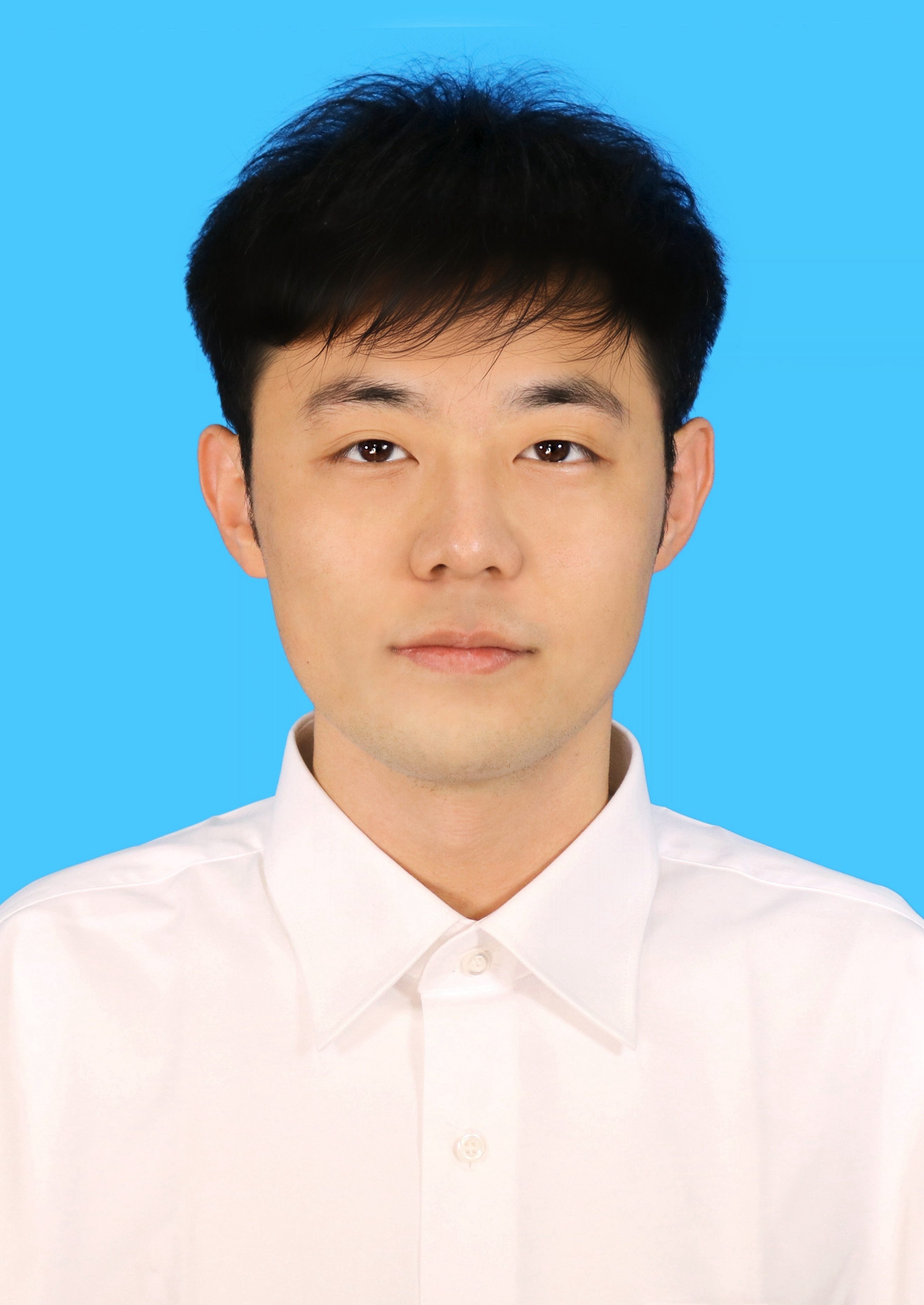}}]{Yize Li}
is currently a Ph.D. student in the School of Computer Science at Hangzhou Dianzi University. His research primarily focuses on visualization and human–computer interaction for large language model applications. His work explores visual analytics, interactive techniques, and human-AI collaboration to support understanding, analysis, and generation in natural language processing tasks.
\end{IEEEbiography}

\begin{IEEEbiography}[{\includegraphics[width=1in,height=1.25in,clip,keepaspectratio]{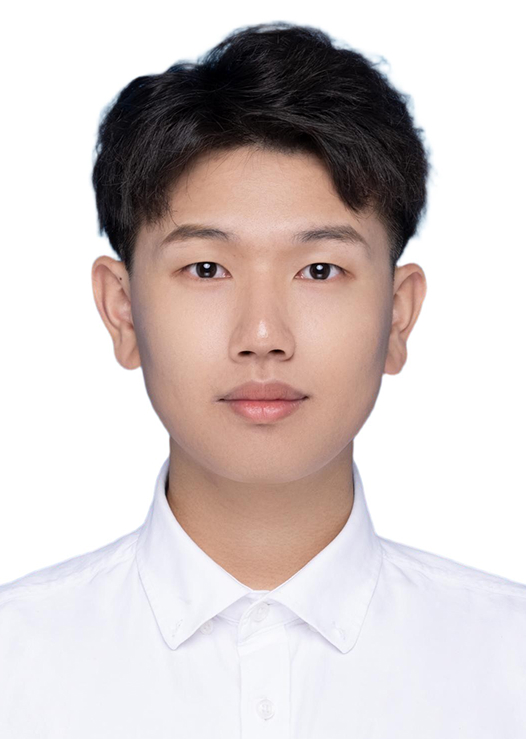}}]{Ruiqi Yu}
is currently a master's degree candidate in the School of Computer Science at Hangzhou Dianzi University. His research interests include data visualization, human–computer interaction, and intelligent education, with a focus on multimodal visual analytics and interactive techniques for visual content understanding and generation.
\end{IEEEbiography}

\begin{IEEEbiography}[{\includegraphics[width=1in,height=1.25in,clip,keepaspectratio]{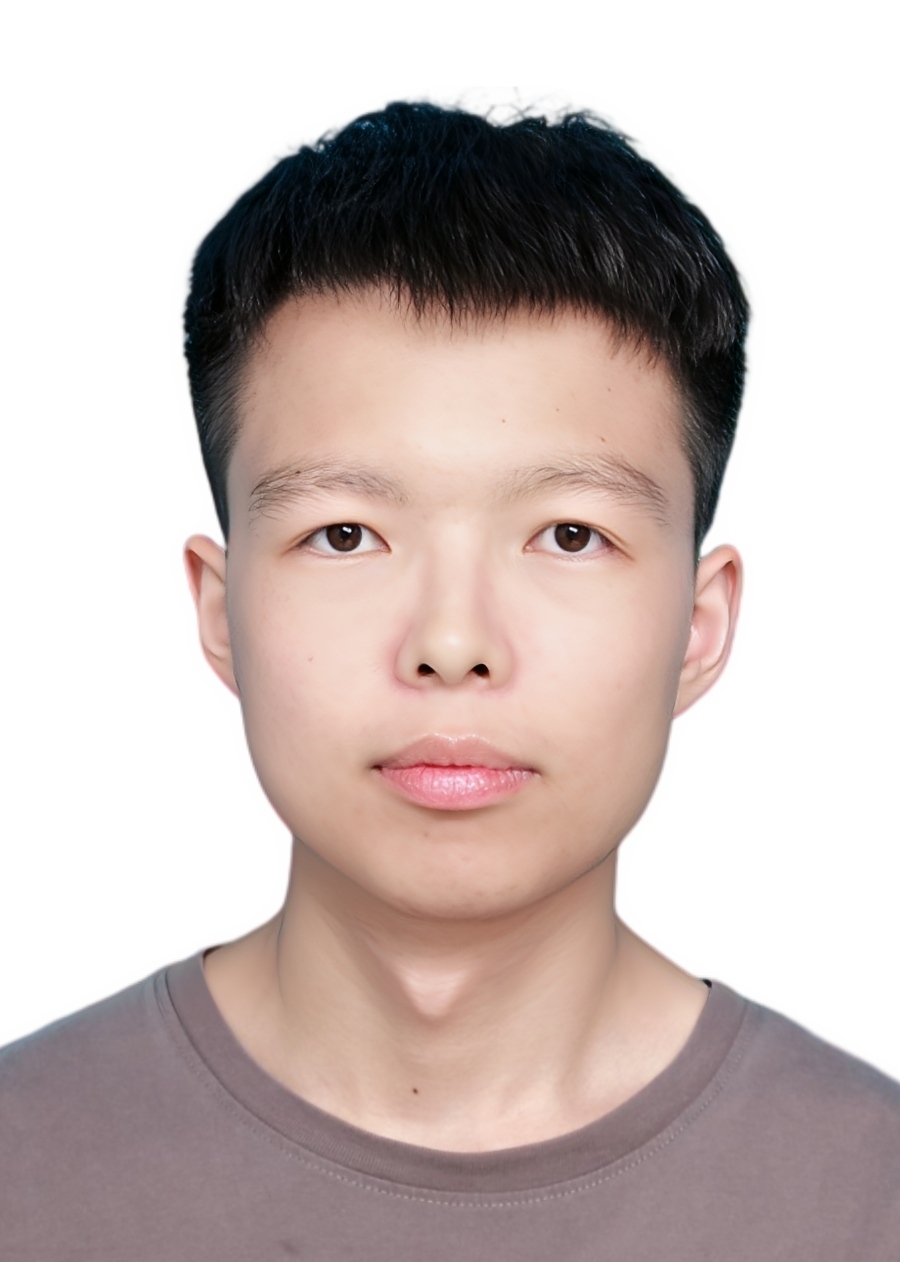}}]{Tianya Pan} 
is currently a master's degree candidate in the School of Computer Science at Hangzhou Dianzi University. His research interests include data visualization, human--computer interaction, and intelligent agents, with a focus on visual analytics for understanding, controlling, and evaluating agent-based systems.
\end{IEEEbiography}

\begin{IEEEbiography}[{\includegraphics[width=1in,height=1.25in,clip,keepaspectratio]{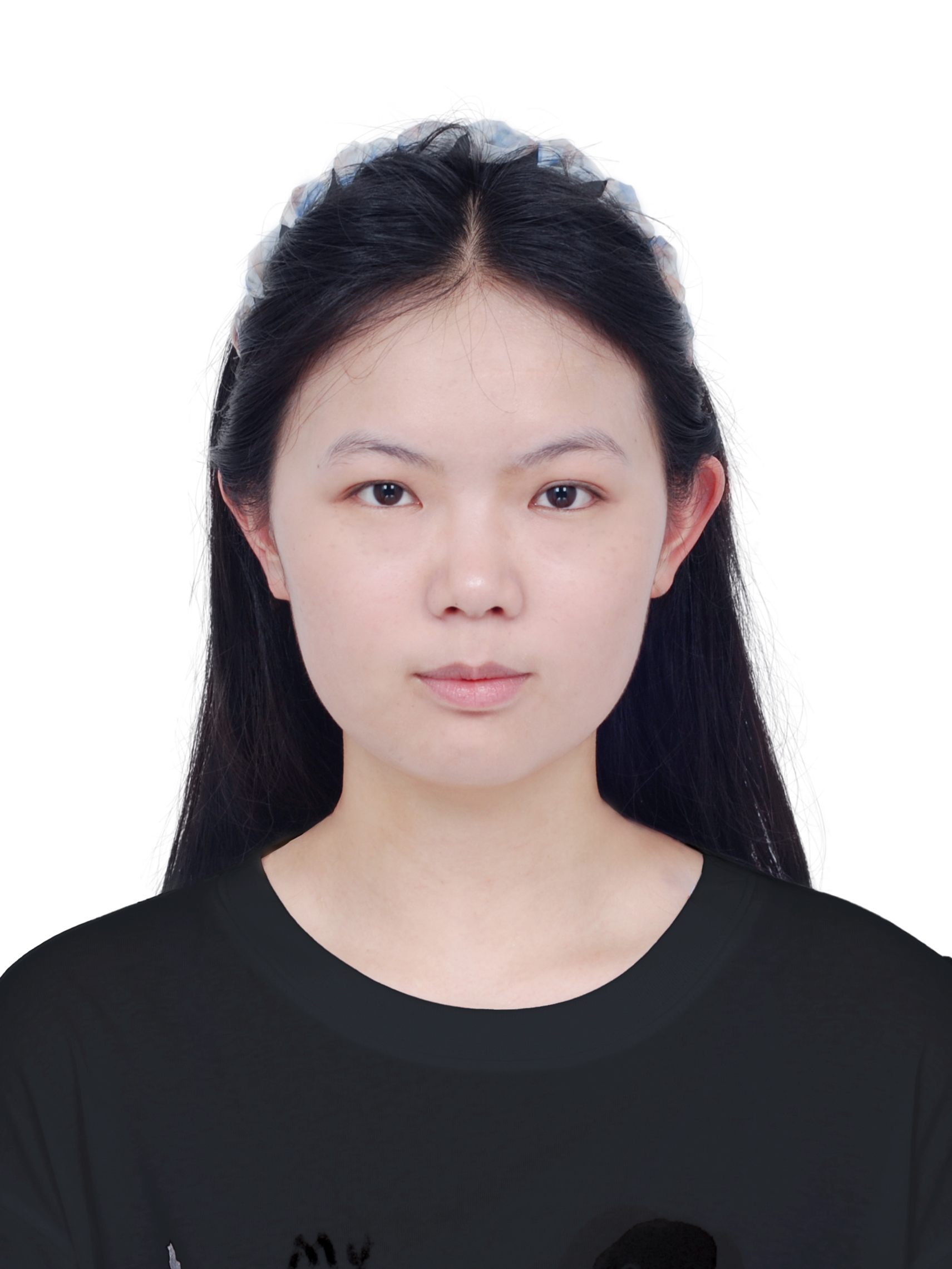}}]{Ningxin Li}
is currently a master's degree candidate in the School of Computer Science at Hangzhou Dianzi University. Her research interests include data visualization, human--computer interaction, and spatial data analysis, with a focus on spatial visualization and visual analytics techniques for exploring and understanding complex spatial and spatiotemporal data.
\end{IEEEbiography}

\begin{IEEEbiography}[{\includegraphics[width=1in,height=1.25in,clip,keepaspectratio]{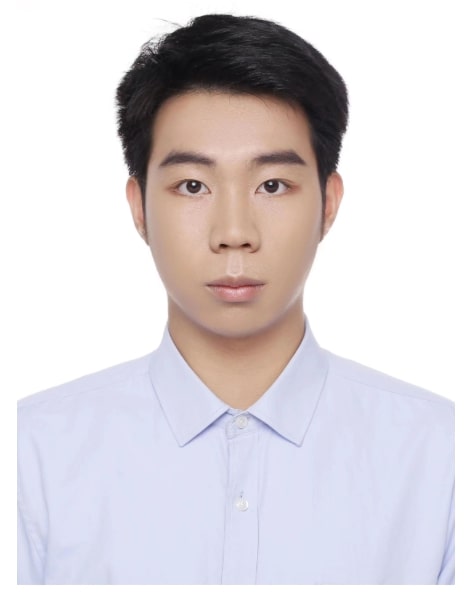}}]{Songyue Li}
is currently a Ph.D. student in the School of Computer Science at Hangzhou Dianzi University. His research primarily focuses on visualization and human–computer interaction for large language model applications. His work explores visual analytics, interactive techniques, and human-AI collaboration to support data understanding, insight discovery, and intelligent decision-making with AI agents.
\end{IEEEbiography}

\begin{IEEEbiography}[{\includegraphics[width=1in,height=1.25in,clip,keepaspectratio]{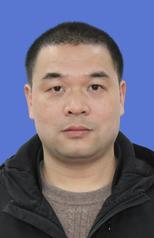}}]{Xiangyang Wu}
is currently a professor in the School of Computer Science at Hangzhou Dianzi University. His research
interests include large-scale data analytics, data visualization and visual analytics, transportation network modeling, etc. He received the Ph.D. degree in math from Zhejiang University, China.
\end{IEEEbiography}

\begin{IEEEbiography}[{\includegraphics[width=1in,height=1.25in,clip,keepaspectratio]{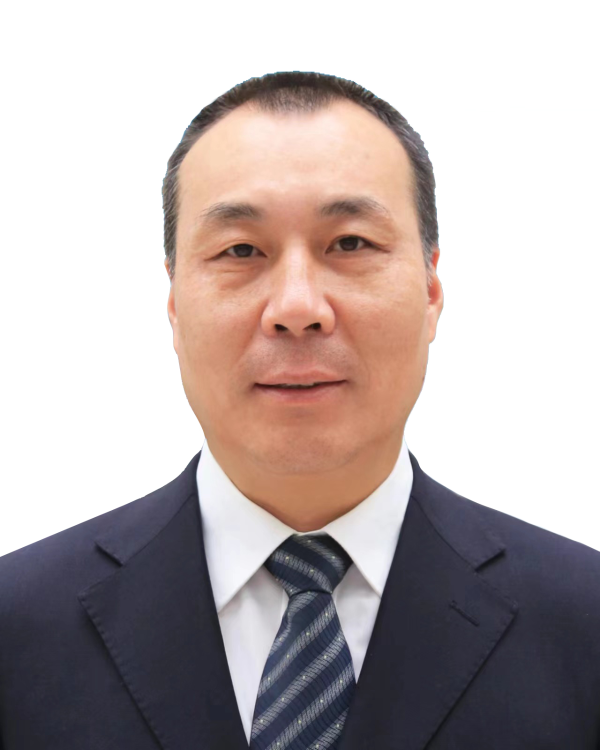}}]{Jinchang Li}
is currently a professor in the School of Data Sciences at Zhejiang University of Finance \& Economics. His research interests include statistical measurement, data quality assessment, and knowledge discovery and applications. He received the Ph.D. degree in economics from Xiamen University, China.
\end{IEEEbiography}

\begin{IEEEbiography}[{\includegraphics[width=1in,height=1.25in,clip,keepaspectratio]{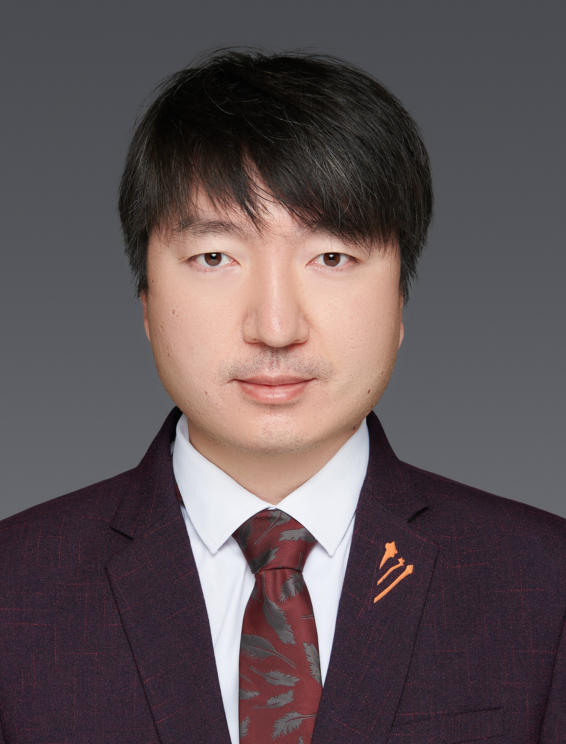}}]{Zhiguang Zhou}
is currently a professor in the School of Computer Science at Hangzhou Dianzi University. His research interests include data visualization, visual analytics and knowledge graph mining. He received his Ph.D. in Computer Science from the State Key Laboratory of CAD\&CG at Zhejiang University. 
\end{IEEEbiography}

\vfill

\end{document}